\documentclass[fleqn,usenatbib,useAMS]{mnras}

\usepackage{graphicx}	
\usepackage{amsmath}	
\usepackage{amssymb}	
\usepackage{multicol}        
\usepackage{bm}		
\usepackage{pdflscape}	

\usepackage{xcolor}
\usepackage{color}
\usepackage{graphicx}
\usepackage{dcolumn}
\usepackage{bm}
\usepackage{hyperref}
\usepackage{amsmath}
\usepackage{multirow}
\usepackage{cancel}

\newcommand{\Ms}{{\rm ~M}_\odot}

\newcommand{\yrgpc}{{\rm ~yr}^{-1}{\rm ~Gpc}^{-3}}

\newcommand{\gw}{{\rm GW}}
\newcommand{\der}{{\rm d}}
\newcommand{\bbh}{{\rm BBH}}
\newcommand{\ham}{{\rm HM}}

\newcommand{\nbsix}{\textsc{Nbody6++GPU~}}

\newcommand{\bse}{\textsc{BSE~}}
\newcommand{\mcl}{\textsc{McLuster~}}

\newcommand{\dragonii}{\textsc{Dragon-II~}}

\usepackage[T1]{fontenc}
\usepackage{ae,aecompl}

\title[The \textsc{Dragon-II} simulations - III.]{The \textsc{Dragon-II} simulations - III. Compact binary mergers in clusters with up to 1 million stars: mass, spin, eccentricity, merger rate and pair instability supernovae rate.}
\author[M. Arca Sedda et al]{Manuel Arca Sedda$^{1,2,3}$%
\thanks{Contact e-mail:\href{mailto:m.arcasedda@gmail.com}{manuel.arcasedda@gssi.it}},
Albrecht W. H. Kamlah$^{4,3}$, Rainer Spurzem$^{3,5,6}$,
\newauthor
Francesco Paolo Rizzuto$^{7}$, Mirek Giersz$^{9}$, Thorsten Naab$^{7}$, Peter Berczik$^{3,10,11}$
\\
$^{1}$ Gran Sasso Science Institute (GSSI), 67100 L’Aquila, Italy\\
$^{2}$ Physics and Astronomy Department Galileo Galilei, University of Padova, Vicolo dell'Osservatorio 3, I--35122, Padova, Italy\\
$^{3}$ Astronomisches Rechen-Institut, Zentrum f\"{u}r Astronomie der Universit\"{a}t  Heidelberg, M\"onchhofstr. 12-14, D-69120 Heidelberg, Germany\\
$^{4}$ Max-Planck-Institut f\"ur Astronomie, K\"onigstuhl 17, 69117 Heidelberg, Germany \\
$^{5}$ National Astronomical Observatories and Key Laboratory of Computational Astrophysics, Chinese Academy of Sciences, 20A Datun Rd.,Chaoyang District, 100101, Beijing, China\\
$^{6}$ Kavli Institute for Astronomy and Astrophysics, Peking University, Yiheyuan Lu 5, Haidian Qu, 100871, Beijing, China\\
$^{7}$ Department of Physics, University of Helsinki, Gustaf Hällströmin katu 2, FI-00014, Helsinki, Finland\\
$^{8}$ Max Planck Institute for Astrophysics, Karl-Schwarzschild-Str. 1, D-85740, Garching, Germany\\
$^{9}$ Nicolaus Copernicus Astronomical Centre, Polish Academy of Sciences, ul. Bartycka 18, 00-716 Warsaw, Poland\\
$^{10}$ Konkoly Observatory, Research Centre for Astronomy and Earth Sciences, E\"otv\"os Lor\'and Research Network (ELKH), MTA Centre of Excellence, Konkoly Thege Mikl\'os \'ut 15-17, 1121 Budapest, Hungary \\
$^{11}$ Main Astronomical Observatory, National Academy of Sciences of Ukraine, 27 Akademika Zabolotnoho St., 03680, Kyiv, Ukraine \\
}

\date{\today}

\pubyear{2022}

\begin{document}
\label{firstpage}
\pagerange{\pageref{firstpage}--\pageref{lastpage}}
\maketitle

\begin{abstract}
Compact binary mergers forming in star clusters may exhibit distinctive features that can be used to identify them among observed gravitational-wave (GW) sources. Such features likely depend on the host cluster structure and the physics of massive star evolution. Here, we dissect the population of compact binary mergers in the \textsc{Dragon-II} simulation database, a suite of 19 direct $N$-body models representing dense star clusters with up to $10^6$ stars and $<33\%$ of stars in primordial binaries. We find a substantial population of black hole binary (BBH) mergers, some of them involving an intermediate-mass BH (IMBH), and a handful mergers involving a stellar BH and either a neutron star (NS) or a white dwarf (WD). Primordial binary mergers, $\sim 30\%$ of the whole population, dominate ejected mergers. Dynamical mergers, instead, dominate the population of in-cluster mergers and are systematically heavier than primordial ones. Around $20\%$ of \textsc{Dragon-II} mergers are eccentric in the LISA band and $5\%$ in the LIGO band. We infer a mean cosmic merger rate of $\mathcal{R}\sim 12(4.4)(1.2)$ yr$^{-1}$ Gpc$^3$ for BBHs, NS-BH, and WD-BH binary mergers, respectively, and discuss the prospects for multimessenger detection of WD-BH binaries with LISA. We model the rate of pair-instability supernovae (PISNe) in star clusters and find that surveys with a limiting magnitude $m_{\rm bol}=25$ can detect $\sim 1-15$ yr$^{-1}$ PISNe. Comparing these estimates with future observations could help to pin down the impact of massive star evolution on the mass spectrum of compact stellar objects in star clusters.
\end{abstract}

\begin{keywords}
methods: numerical – galaxies: star clusters: general – stars: general, black holes
\end{keywords}



\section{Introduction}
In less than a decade, the LIGO-Virgo-Kagra (LVK) collaboration discovered 76 confident gravitational-wave (GW) sources associated to merging stellar black holes (BHs) and neutron stars (NSs) \citep{2021arXiv211103634T}. This number raises up to 90 if one considers the population of events with a probability to have an astrophysical origin $> 0.5$ \citep{2021arXiv211103606T}, and it is destined to further increase by the end of the fourth observation run. Measurable quantities like component masses, spins, or the orbital eccentricity, and the merger rate of different types of compact binary mergers can represent the keys to identify the signatures of different formation channels \citep{2019MNRAS.482.2991A,2020ApJ...894..133A,2020ApJ...899L...1Z,2021arXiv210912119A,2021MNRAS.507.5224B,2022MNRAS.511.5797M}. From the theoretical standpoint, there is a plethora of mechanisms proposed to explain the formation of compact binary mergers, like isolated binary evolution \citep{2002ApJ...572..407B,2012ApJ...759...52D,2016A&A...594A..97B,2018MNRAS.474.2959G,2019MNRAS.485..889S}, dynamical pairing in dense star clusters \citep{2002ApJ...576..894M,2010MNRAS.407.1946D, 2016PhRvD..93h4029R, 2017MNRAS.464L..36A,2018MNRAS.473..909B,2019MNRAS.487.2947D,2022MNRAS.512..884R}, formation in AGN disks \citep{2012MNRAS.425..460M,2017MNRAS.464..946S, 2018ApJ...866...66M,2020ApJ...898...25T}, secular dynamics involving three compact objects \citep{2018MNRAS.480L..58A,2021ApJ...907L..19V} or a binary orbiting a supermassive black hole \citep{2012ApJ...757...27A,2018ApJ...856..140H,2019MNRAS.488...47F,2020ApJ...891...47A}, and primordial BH evolution \citep{1974MNRAS.168..399C,2016PhRvD..94h3504C,2016PhRvL.117f1101S}. The majority of the aforementioned mechanisms relies on the assumption that compact objects are the relic of massive stars, and therefore they suffer the uncertainties affecting stellar evolution. 

For example, the insurgence of pair instability supernova (PISN) and pulsational pair instability supernova (PPISN) mechanisms can carve in the BH mass spectrum the so-called upper-mass gap, a region extending in the range $m_{\rm gap} = 40-150\Ms$ where no remnants are expected. The boundaries of the gap are highly uncertain and depend on many poorly constrained quantities, like stellar rotation, rate of nuclear reactions, stellar evolution model \citep{2021ApJ...912L..31W,2021MNRAS.504..146V,2019ApJ...882..121S,2019ApJ...887...53F,2021MNRAS.501.4514C, 2022arXiv221111774I}. The presence of several upper mass-gap BH candidates in the LVK source catalogue poses the question about the origin of these BHs. Stellar mergers, star-BH interactions, and repeated BH mergers represent possible pathways to overcome (P)PISN \citep[e.g.][]{2019MNRAS.485..889S,2022A&A...665A..20B, 2022MNRAS.516.1072C,2022arXiv220403493B} and produce merging compact objects in dense star clusters \citep[e.g.][]{2018PhRvL.120o1101R,2020MNRAS.497.1043D, 2020ApJ...903...45K,2021MNRAS.507.5132D, 2021MNRAS.501.5257R,2021ApJ...920..128A,2022MNRAS.512..884R}.

Spins could carry crucial information on the BH formation scenario and help placing constraints on the evolution of massive stars, but little is known about the distribution of stellar BH natal spins. Observations of merging BHs indicate that the spin distribution follows a Maxwellian distribution, with a peak around $\chi_{\rm BH} \sim 0.2-0.5$ \citep{2021arXiv211103634T}. However, stellar BHs detected in low-mass X-ray binaries (LMXBs) are characterised by spins broadly distributed in the whole allowed range \citep{2015ApJ...800...17F}, whilst those in high-mass X-ray binaries (HMXBs) involve BHs almost maximally spinning \citep[see e.g.][]{2019ApJ...870L..18Q,reynolds21}. 
Despite these differences may suffer observation biases, they may represent peculiarities of different evolutionary pathways. 
Efficient angular momentum transport driven by magnetic stars could trigger the formation of BHs with natal spins as small as $\chi_{\rm BH} \lesssim 0.01$, a mechanism proposed for BHs forming from single stars and in binaries with a negligible mass transfer among the components \citep{2019ApJ...881L...1F}. Significant mass transfer, instead, has been proposed to produce BHs with spin in a broad range in LMXB, even for BHs spinless at birth \citep{2015ApJ...800...17F}, and nearly extremal BHs in HMXBs \citep{2019ApJ...870L..18Q,2022ApJ...938L..19G}. 
Common envelope evolution in massive stellar binaries can lead  to merging BBHs consisting in a nearly non-rotating BH \citep{2018A&A...616A..28Q, 2020A&A...635A..97B}, although this strongly depends on the stellar evolution adopted \citep{2020A&A...636A.104B}, and a BH companion with a spin spanning the whole allowed range of values  \citep{2018A&A...616A..28Q,2020A&A...635A..97B,2020A&A...636A.104B}.

Amplitude aside, also the alignment of the spin vectors among themselves and with the binary angular momentum can affect both the waveform, the final merger remnant mass and spin, and the recoil kick (e.g. see Equation \ref{eqKick2}). From an "observational" perspective, measuring the spin components is intrinsically hard and their directions generally vary owing to precession, thus the spin of observed mergers can be characterised through the so-called effective spin parameter \citep{2008PhRvD..78d4021R,2010PhRvD..82f4016S,2011PhRvL.106x1101A}
\begin{equation}
\chi_{\rm eff} = \frac{\vec{\chi}_1 + q\vec{\chi}_2}{1+q}\cdot \vec{L},
\end{equation} 
where $q < 1$ is the binary mass ratio, $\vec{\chi}_{1,2}$ are the two component spin vectors and $\vec{L}$ is the binary orbital angular momentum.
Observations of BBH mergers suggest that $\chi_{\rm eff}$ may increase at increasing the binary merger mass ratio, although some merging binaries exhibit a negative value of $\chi_{\rm eff}$ \citep{2021arXiv211103634T}, a feature generally associated with dynamical sources.

The orbital eccentricity at merger could represent another distinguishing feature of compact binary mergers, as dynamical interactions could trigger the formation of fairly eccentric ($>0.1$) sources contrarily to mergers forming from isolated binaries \citep[see e.g.][]{2016PhRvD..94f4020N}. It has been recently claimed that up to four LVK sources may be eccentric \citep{2019MNRAS.490.5210R,2020ApJ...903L...5R,2022NatAs...6..344G,2022ApJ...940..171R}, although the effects of eccentricity and precession can lead to degeneracies in GW data analysis, making the eccentricity a poorly constrained quantity \citep[see e.g.][]{2023MNRAS.519.5352R}.

Alongside GWs, the detection of (P)PISNe can represent a key piece to understand the final stages of massive stars' life. So far, only a few, most of which controversial, PISN and PPISN candidates have been observed in the last two decades \citep{2009Natur.462..624G, 2019ApJ...881...87G, 2022ApJ...938...57W}. The rarity of PISNe observations sets an intrinsic limit on the frequency of PISNe in star clusters, a quantity poorly constrained in theoretical and numerical models.

Dynamical interactions among stars in dense and massive star clusters can trigger both the formation of merging binaries and the development of PISNe, either from single massive stars or from stellar merger products. Young and intermediate-age star clusters are particularly interesting environments where these sources can form, because they are still in their dynamical youth, when cluster mass-loss and expansion did not yet affected substantially the cluster structure and the interaction rate among stars is maximal. There is vast literature investigating the formation and evolution of merging BHs in star clusters via different techniques, e.g. direct $N$-body simulations \citep{2010MNRAS.402..371B,2010MNRAS.407.1946D,2019MNRAS.487.2947D,2020MNRAS.497.1563R, 2020CmPhy...3...43A, 2021MNRAS.507.5132D, 2018MNRAS.473..909B,2021MNRAS.500.3002B,2022A&A...665A..20B,2022MNRAS.509.4713W,2022MNRAS.513.4527C,2022MNRAS.512..884R}, Monte Carlo simulations \citep{2016PhRvD..93h4029R,2017MNRAS.464L..36A,2019PhRvD..99f3003K,2019PhRvD.100d3027R,2020ApJ...888L..10Y,2020ApJ...903...45K,2022MNRAS.514.5879M}, and semi-analytic tools \citep{2018PhRvL.121p1103F, 2019MNRAS.486.5008A, 2019MNRAS.482.2991A, 2020ApJ...894..133A, 2020MNRAS.492.2936A, 2021ApJ...908L..29G, 2021Symm...13.1678M, 2021arXiv210912119A, 2022MNRAS.511.5797M, 2022arXiv220801081A, 2022arXiv221010055K}. However, there is lack of direct $N$-body simulations of particularly dense ($>10^5 ~\Ms$ pc$^{-3}$) and massive ($>100,000\Ms$) star clusters, owing to the computational cost required to simulate such systems. Exploring this range of mass and densities with $N$-body models can complement the already existing simulations and can offer a term of comparison to Monte Carlo simulations (see Figure 1 in Arca Sedda et al 2023a, hereafter AS-I). 

In this work, which represents the third of a series, we present results from the \dragonii star cluster database, a suite of 19 direct $N$-body simulations of young and intermediate-age star clusters comprised of up to 1 million stars and up to $33\%$ of stars initially in binaries, characterised by typical densities $\rho = (1.2\times10^4 - 1.5\times 10^7)\Ms$ pc$^{-3}$. 

In our previous papers, we focused on the general properties of our cluster models and their compact object populations (paper AS-I) and the processes that regulate the formation and growth of IMBHs (Arca Sedda et al 2023b, hereafter AS-II).

Here, we dissect the properties of BH-BH, BH-NS, and BH-WD mergers developing in the \dragonii star cluster database, a suite of 19 direct $N$-body simulations of star clusters comprised of up to 1 million stars and up to $33\%$ of stars initially in binaries (details about these models are discussed in our companion paper AS-I), performed with the \nbsix code\footnote{\url{https://github.com/nbody6ppgpu/Nbody6PPGPU-beijing}}. The paper is organised as follows: in Section \ref{sec:meth} we briefly summarise the main features of our models; Section \ref{sec:res} discusses the main properties of compact binary mergers in our models, focusing on the component masses and mass ratios, the eccentricity at merger, and the possible signatures that can identify their formation history; in Section \ref{sec:disc} we explore the impact of BH natal spins onto the global properties of the population, and we adopt a cosmologically motivated framework to infer the compact binary merger rate, the detection perspectives for future low-frequency GW detections, and the frequency rate and detection perspectives in magnitude limited surveys of PISNe; Section \ref{sec:end} summarises the main results of this work.

\section{Numerical methods}
\label{sec:meth}

\subsection{The \dragonii clusters}

The \dragonii simulation database consists of 19 star cluster models characterised by an initial number of stars $N = (1.2,~ 3,~ 6,~ 10)\times 10^5$,  half-mass radius $R_\ham = (0.48,~0.80,~1.76)$ pc, and an initial binary fraction $f_b = 0.05-0.2$. In the following, we briefly summarise the main properties of \dragonii clusters, referring the interested readers to our companion paper AS-I for more details on the run properties. 

To initialise the \dragonii clusters we exploit the \mcl tool \citep[][Leveque in prep.]{2011MNRAS.417.2300K, 2022MNRAS.511.4060K, 2022MNRAS.514.5739L}. 

Each cluster is modelled according to a \cite{1966AJ.....71...64K} profile with adimensional potential well $W_0 = 6$. We adopt an initial metallicity $Z = 0.0005$, typical of several clusters possibly hosting a dense sub-system of compact objects or an IMBH, like NGC3201 or NGC6254 \citep{2018MNRAS.478.1844A, 2018MNRAS.479.4652A,2020ApJ...898..162W}.

Star masses are drawn according to a \cite{2001MNRAS.322..231K} initial mass function limited in the range $m_{\rm ZAMS} = (0.08-150)\Ms$. Stars in primordial binaries are paired depending on their mass, with stars heavier than $>5\Ms$ paired according to a flat mass-ratio distribution, and lighter stars paired randomly. Binary eccentricities are distributed according to a thermal distribution, $P(e){\rm d}e = e^2{\rm d}e$, while initial semimajor axes are assigned according to a distribution flat in logarithmic values limited between the sum of stars' radii and a maximum value of $50$ AU.

The host galaxy potential is modelled through a Keplerian potential assuming a total mass of $M_{\rm gal} = 1.78\times10^{11}\Ms$. All \dragonii clusters are placed on a circular orbit around this galaxy model at a distance of $R_{\rm clu} = 13.3$ kpc. The adopted galaxy mass and orbital radius lead to a value of the circular velocity compatible with what is observed in the Milky Way.

The resulting tidal radius is much larger than the cluster half-mass radius. Therefore, \dragonii models are initially underfilling their Roche lobe, which implies that the initial impact of the host galaxy potential is negligible.
 
All simulations are terminated when either the mean BH mass falls below $\langle m_{\rm BH} \rangle \lesssim 15\Ms$, there are no BHs with a mass above $30\Ms$, or the simulated time exceeds at least one relaxation time. As a result, the simulation time in \dragonii models spans a range $T_{\rm sim} = 0.1-2.3$ Gyr, corresponding to $0.8-80$ times the initial half-mass relaxation time (see also Table \ref{tab:t1}).

Over the simulated time, we find a nice overlap (see also Figure 2 in paper AS-I) between the evolution of \dragonii clusters' mass and half-mass radius and observed properties of young and intermediate-age massive clusters in the Milky Way \citep{2010ARA&A..48..431P}, the Magellanic clouds \citep{2021MNRAS.507.3312G}, and other galaxies in the local Universe like Henize 2-10 \citep{2014ApJ...794...34N} or M83 \citep{2015MNRAS.452..525R}. In this sense, \dragonii models can represent one possible evolutionary pathways of (relatively) young massive clusters.

\subsection{The \nbsix code}

\dragonii simulations have been performed with the \nbsix code \citep{2015MNRAS.450.4070W}, a state-of-the-art direct $N$-body integrator that runs on high performance computing hardware equipped with graphic-processing-units  \citep[GPUs,][]{1999JCoAM.109..407S,2012MNRAS.424..545N,2015MNRAS.450.4070W}. The code is part of the famous \textsc{NBODY} code series that was pioneered almost sixty years ago by Sverre Aarseth \citep{1974A&A....37..183A,1999JCoAM.109..407S,1999PASP..111.1333A,2003gnbs.book.....A,2008LNP...760.....A,2012MNRAS.424..545N,2015MNRAS.450.4070W,2022MNRAS.511.4060K}.

The code implements a 4th-order Hermite integrator scheme with adaptive time-step based on the Ahmad-Cohen scheme for neighbours \citep{1973JCoPh..12..389A}, and implements a treatment for close encounters and few-body dynamics via the Kustaanheimo-Stiefel regularisation \citep[][]{Stiefel1965} and chain regularisation \citep[][]{1993CeMDA..57..439M}.

Stellar evolution in \nbsix is based on an upgraded version of the population synthesis code \bse \citep{2002MNRAS.329..897H}. The main features of this state-of-the-art version, named \textsc{BSE++}, are described in detail in \cite{2022MNRAS.511.4060K} \citep[but see also][]{2020A&A...639A..41B}. We adopt the so-called level B of stellar evolution \citep[see][]{2022MNRAS.511.4060K}, whose main characteristics are: delayed supernova (SN) scheme \citep{2012ApJ...749...91F}, pair- and pulsation pair-instability supernova (PISN and PPISN) treated following \cite{2016A&A...594A..97B, 2021MNRAS.500.3002B}, fallback prescription for NS/BH natal kicks, and metallicity-dependend winds for massive stars \cite{2001A&A...369..574V,2010ApJ...714.1217B}. We refer the reader to \cite{2022MNRAS.511.4060K} and paper AS-I for further details.

The common envelope phase in binaries is modelled through the widely known $\alpha_{\rm CE}-\lambda_{\rm CE}$ scheme, which enables us to regulate the fraction of orbital energy injected into the envelope ($\alpha_{\rm CE}$) and to scale the binding energy of the envelope by a factor $\lambda_{\rm CE}$. In this work, we adopt $\alpha_{\rm CE} =  3$ and $\lambda = 0.5$ \citep{2018MNRAS.480.2011G, 2022MNRAS.511.4060K}.

The adopted stellar evolution recipes imply that the stellar BH mass-spectrum in \dragonii clusters is limited to $m_{\rm BH,max} = 40.5\Ms$ \citep{2016A&A...594A..97B}, unless BHs form from stellar mergers or star-BH interactions. In the latter case, \nbsix parametrises the amount of mass accreted in a strong star-BH interaction or collision via an accretion parameter $f_c$ \citep{2021MNRAS.501.5257R,2022MNRAS.512..884R}, which we set to $f_c=0.5$. We refer the reader to \cite{2021MNRAS.501.5257R,2022MNRAS.512..884R} for a discussion about the impact of $f_c$ on BH evolution.

\subsubsection{Modelling the final stages of compact object binary mergers}
The dynamics of relativistic binaries is followed via the orbit-average formalism \citep{1964PhRv..136.1224P}, which enables us to follow the evolution of compact binaries and their coalescence inside the cluster, similarly to previous works \citep[see e.g. ][]{2019MNRAS.487.2947D, 2020MNRAS.497.1043D,2020MNRAS.498..495D,2021MNRAS.507.5132D,2021MNRAS.501.5257R,2022MNRAS.512..884R,2021MNRAS.507.3612R,2022MNRAS.517.2953T}.
 
In its current implementation, \nbsix follows the dynamics of relativistic binaries also if they are part of triples \citep[see e.g.][]{2021MNRAS.501.5257R} and multiple systems, as well as if they form via hyperbolic interactions. 

However, the BBH evolution is not followed down to the merger, rather the binary is decoupled from dynamics and promtply merged when the BBH pericentre falls below a critical value, which we set to $10^2$ Schwarzschild radii, i.e. $a_{\rm dec} = 2kGm_{\rm bin}/c^2 = kR_{\rm Sch}$ with $k=100$. 

Adopting such limiting separation ensures that the binary is unlikely to undergo any further interaction with surrounding stars before merging. Considering the range of binary masses ($1-300\Ms$), star cluster masses ($<10^6\Ms$) and half-mass radii ($0.1-3$ pc) explored in this work, it is easy to show that the binary---single interaction timescale $t_{2-1} = (n \sigma \Sigma)^{-1}$ -- with $n$ the cluster density, $\sigma$ the cluster velocity dispersion, and $\Sigma$ the binary cross section -- is generally $>10^8$ larger than the binary inspiral timescale, $t_{\rm insp} \propto a^4/(m_1m_2m_{\rm bin})$ \citep{1963PhRv..131..435P}. 
Moreover, the typical merger time for a binary with mass $m_{\rm bin} < 200 \Ms$ and separation $a_{\rm dec}$ is generally $t_{\rm insp} < 100$ yr, i.e. much smaller than the cluster crossing time, $t_{\rm step}\sim 10^5$ yr. 

Therefore, our procedure ensures reliabile results while reducing the computational effort required to simulate the evolution of a binary with an orbital period of minutes or hours.

The pre-merger stages of the merging binary orbits are reconstructed by retrieving the orbital parameters at decoupling and integrating the orbit via the \cite{1964PhRv..136.1224P} equations:
\begin{eqnarray}
\frac{\der a}{\der t} &=& -\frac{64}{5}\beta(m_{1},m_{2})\frac{F(e)}{a^3},\\
\frac{\der e}{\der t} &=& -\frac{304}{15}\beta(m_{1},m_{2})\frac{e G(e)}{a^4},
\label{peters2}
\end{eqnarray}
with
\begin{eqnarray}
F(e)          &=& (1 - e^2)^{-7/2}\left(1 +\frac{73}{24} e^2 + \frac{37}{96}e^4\right);\\
\beta(m_{1},m_{2})  &=& (G^3/c^5) m_{1}m_{2}(m_{1}+m_{2});\\
G(e)          &=& (1-e^2)^{-5/2}\left(1+\frac{121}{304}e^2\right).
\end{eqnarray}
Along with the orbital evolution we calculate the associated GW strain and frequency \cite[see e.g.][]{1963PhRv..131..435P,2012ApJ...752...67K,2021A&A...650A.189A}.

Natal spins of stellar BHs can be assigned according to different distribution, three of which are based on physical stellar model, namely the ``Geneva'', ``MESA'', and ``Fuller'' models \citep[see e.g.][]{2022MNRAS.511.4060K}, and four are rather generic, namely zero-spins, uniform spin distribution, Gaussian spin distribution with mean value $\chi = 0.5$ and dispersion $\sigma_\chi = 0.2$, and Maxwellian distribution with dispersion $\sigma_\chi = 0.2$.

In this work, whenever spins are taken into account during the simulation we assume a Gaussian distribution with $\chi = 0.5$ for stellar BHs, whilst for IMBHs we decide on a case by case basis, depending on the IMBH formation scenario (see paper AS-II).

Compact binary merger products are assigned a final mass and spin calculated via numerical relativity fitting formulae \citep{2017PhRvD..95f4024J,2020ApJ...894..133A} and a relativistic recoil, generated by asymmetric GW emission \citep{2007PhRvL..98w1102C,2008PhRvD..77d4028L,2012PhRvD..85h4015L}, expressed via the following relation:
\begin{eqnarray}
\vec{v}_\gw   =& v_m\hat{e}_{\bot,1} + v_\bot(\cos \xi \hat{e}_{\bot,1} + \sin \xi \hat{e}_{\bot,2}) + v_\parallel \hat{e}_\parallel, \label{eqKick1}\\
v_m         =& A\eta^2 \sqrt{1-4\eta} (1+B\eta), \\
v_\bot      =& \displaystyle{\frac{H\eta^2}{1+q_\bbh}}\left(S_{2,\parallel} - q_\bbh S_{1,\parallel} \right), \\
v_\parallel =& \displaystyle{\frac{16\eta^2}{1+q_\bbh}}\left[ V_{11} + V_A \Xi_\parallel + V_B \Xi_\parallel^2 + V_C \Xi_\parallel^3 \right] \times \nonumber \\
             & \times \left| \vec{S}_{2,\bot} - q_\bbh\vec{S}_{1,\bot} \right| \cos(\phi_\Delta - \phi_1).  \label{eqKick2}
\end{eqnarray}
Here, $\eta \equiv q_\bbh/(1+q_\bbh)^2$ is the symmetric mass ratio, $\vec{\Xi} \equiv 2(\vec{S}_2 + q_\bbh^2 \vec{S}_1) / (1 + q_\bbh)^2$, and the subscripts $\bot$ and $\parallel$ mark the perpendicular and parallel directions of the BH spin vector ($\vec{S}$) with respect to the direction of the binary angular momentum. We assume $A = 1.2 \times 10^4$ km s$^{-1}$, $B = -0.93$, $H = 6.9\times 10^3$ km s$^{-1}$, and $\xi = 145^\circ$ \cite{2007PhRvL..98i1101G,2008PhRvD..77d4028L}, $V_{11} = 3677.76$ km s$^{-1}$, and $V_{A,B,C} = (2.481, 1.793, 1.507)\times 10^3$ km s$^{-1}$. The vector $\vec{\Delta}$ is defined as $\vec{\Delta} \equiv (M_a+M_b)^2 (\vec{S}_b - q_\bbh \vec{S}_a)/(1+q_\bbh)$. The angle between the direction of the infall at merger and the in-plane component of $\vec{\Delta}$, i.e. $\phi_\Delta$, is drawn from a uniform distribution, while $\phi_1 = 0-2\pi$, which represents the phase of the binary, is extracted between the two limiting values according to a uniform distribution. 
In \nbsix, the user can decide to set the GW recoil to zero or to a fixed value, or to calculate it self-consistently via Eqs. \ref{eqKick1} and \ref{eqKick2}, in which case the kick is assigned to the remnant and the resulting energy correction is included in a similar way as it is done for natal BH kicks. 

As described in detail in paper AS-II, in this paper series we adopt a simplified approach to investigate the impact of GW recoil in the simulations, owing to the fact that the relatively small sample of mergers does not enable us to filter out the inevitable stochastic effect of the BH spin directions and amplitudes on the kick amplitude.
The approach consists of three steps. First, we run all simulations assuming no GW recoil. Second, for each merger event in each simulation we evaluate the GR recoil assuming different distribution for BHs natal spins and we determine whether the remnant is likely to be retained or not in the cluster. Third, if a BH undergoes $n$ mergers in a simulation with zero GW kick, we restart the simulation shortly before the $n$-th merger event and enable GW kicks assuming a spin for the merging components that depends on the BH formation history. This enables us to verify whether the remnant can be retained in the cluster and eventually merge again in a $n+1$th merger generation. In paper AS-II, we have shown that this approach permits us to highlight the fact that even when GW kicks are not taken into account, Newtonian dynamics is sufficient to eject all BH remnants from the parent cluster via strong binary-single encounters.

We note that none of the mergers with component masses $<100\Ms$ undergo multiple mergers. This suggests that even adopting zero GW recoils may have a negligible impact on the formation of compact binary mergers with mass $<100 \Ms$.

\section{Results}
\label{sec:res}

\subsection{The population of black hole binary mergers in \textsc{Dragon-II} clusters}
\label{sec:cob}

In this section we describe the main results of our simulations, focusing on the population of compact binary mergers. Table \ref{tab:t1} summarizes the main properties of \dragonii clusters and their compact objects. 
 
\begin{figure}
\includegraphics[width=\columnwidth]{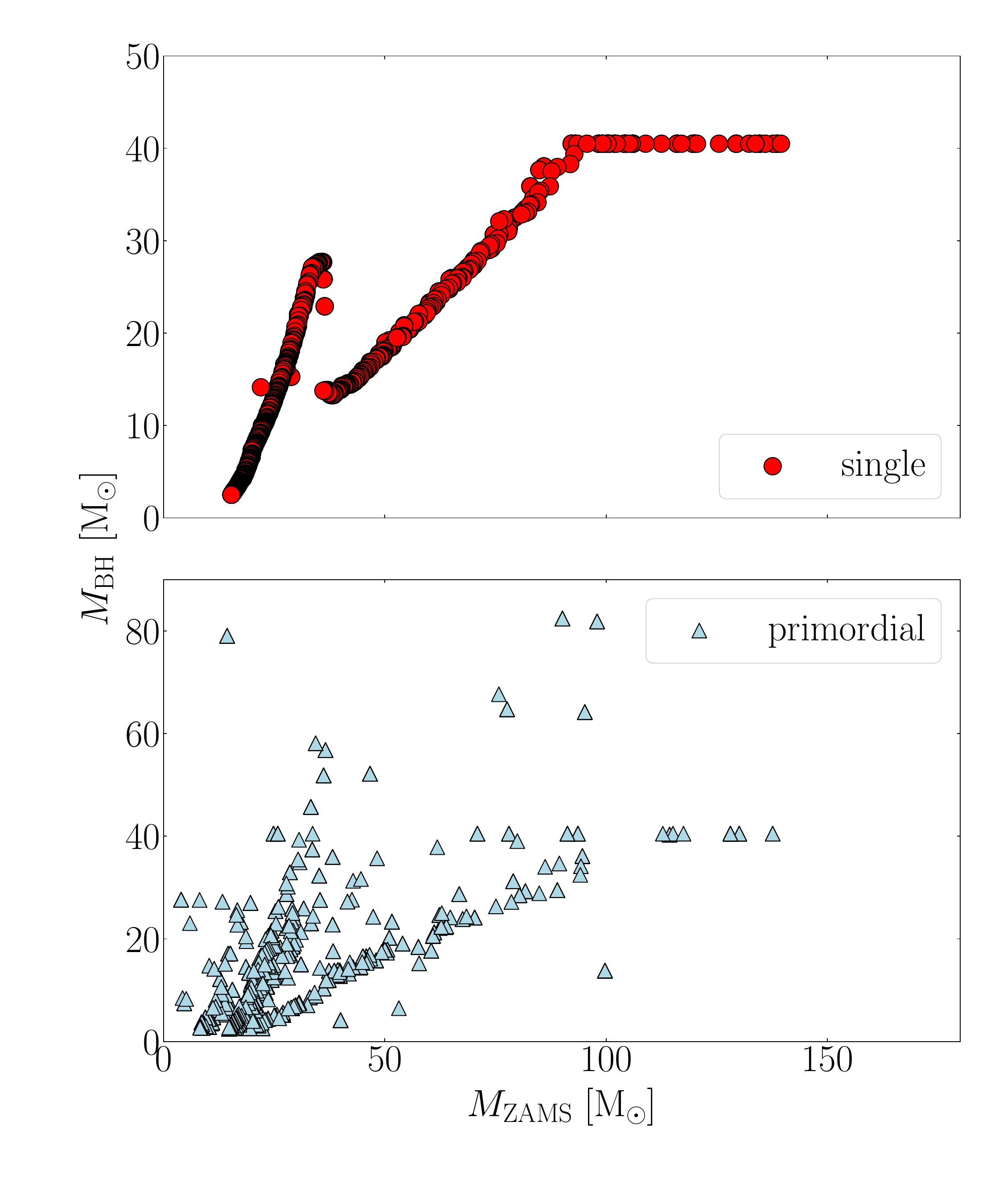}
\caption{Zero age main sequence mass and final mass of stellar BH progenitors. We dissect the population into progenitors that are initially single (top panel, red circles) and those that are in a primordial binary (bottom panel, light blue triangles).}
\label{IFMR}
\end{figure}

The population of BHs formed in \dragonii models and, in general, in star clusters likely suffers both the effects of single and binary stellar evolution and stellar dynamics. To highlight this aspect we show in Figure \ref{IFMR}, for the models with $R_\ham = 0.8$pc and $N=300$k, the so-called initial to final mass relation (IFMR) that links the masses of compact objects and their stellar progenitors. The plot is dissected into BHs with a progenitor initially single or in a primordial binary system. 

The population of BHs forming from single stars generally follows the expectations of the adopted stellar evolution recipes \citep[e.g. see Figure B1 in][]{2022MNRAS.511.4060K}. Deviation from the general trend owes to initially single stars that got caught in a pair and underwent mass-transfer. 
The IFMR of BHs formed from stars in primordial binaries is more complex, being characterised, for example, by BHs in the upper mass-gap with masses in the range $40.5 - 80\Ms$. This highlights the crucial role of binary stellar evolution and dynamics in sculpting the population of BHs in star clusters \citep[e.g., see also Figure 2 in][]{2019MNRAS.487.2947D}.

\begin{table*}
    \centering
    \begin{tabular}{ccccccc|cc|cc|cc|cc|cc|cc|cc}
    \hline
    \hline
        $N_*$ & $M_c$ & $R_h$ & $f_b$ &  $N_{\rm sim}$ & $T_{\rm rlx}$ &$T_{\rm seg}$ & \multicolumn{2}{|c}{$T_{\rm sim}$} & \multicolumn{2}{|c}{$N_{\rm GW, in}$}& \multicolumn{2}{|c}{$N_{\rm GW, out}$} & \multicolumn{2}{|c}{$M_{\rm max}$} & \multicolumn{2}{|c}{$M_{\rm max,fin}$} & \multicolumn{2}{|c}{$N_{>30}$} & \multicolumn{2}{|c}{$N_{>40}$} \\
        $10^3$ & $10^5\Ms$ & pc & &  & Myr & Myr & \multicolumn{2}{|c}{Myr} & \multicolumn{2}{|c}{}&\multicolumn{2}{|c}{}& \multicolumn{2}{|c}{$\Ms$} & \multicolumn{2}{|c}{$\Ms$}& \multicolumn{2}{|c}{}& \multicolumn{2}{|c}{}\\
    \hline
    120 & 0.7 & 1.75 & 0.05& 2& 99 & 2.1 & 2379 & 2326 & 0 & 2 & 2& 0&  64 &   76 & 25 &  34 &   0 &   2 &  0 &  0 \\
    300 & 1.8 & 1.75 & 0.05& 2& 142 & 2.7 & 1196 & 1422 & 0 & 2 & 2& 2&  69 &   77 & 40 &  40 &  13 &  13 &  5 &  1 \\
    1000& 5.9 & 1.75 & 0.05& 2& 233 & 3.4 &  207 &  194 & 1 & 1 & 4& 4&  81 &  146 & 52 &  70 & 149 & 169 & 72 & 85 \\
    120 & 0.7 & 1.75 & 0.2 & 2& 99 & 2.1 & 1710 & 1540 & 2 & 2 & 0& 2& 232 &   81 & 38 &  28 &   2 &   0 &  0 &  0 \\
    300 & 1.7 & 1.75 & 0.2 & 2& 142 & 2.7 &  519 &  793 & 1 & 0 & 7& 5&  92 &   77 & 65 &  47 &  26 &  26 &  8 & 14 \\
    600 & 3.5 & 1.75 & 0.2 & 2& 189 & 3.4 &  205 &  126 & 0 & 0 & 2& 5&  87 &  144 & 59 &  84 &  95 & 103 & 45 & 65 \\
    120 & 0.7 & 0.80 & 0.2 & 2& 30 & 0.7 & 1154 & 1201 & 4 & 3 & 4& 2& 120 &  132 & 21 &  27 &   0 &   0 &  0 &  0 \\
    300 & 1.7 & 0.80 & 0.2 & 2& 44 & 0.8 &  307 &  309 & 1 & 0 & 1& 0&  93 &  107 & 40 &  43 &  15 &  11 &  2 &  2 \\
    120 & 0.7 & 0.47 & 0.2 & 2& 14 & 0.3 & 1149 &  530 & 2 & 2 & 3& 1& 350 &   92 & 50 &  30 &   1 &   0 &  1 &  0 \\
    300 & 1.7 & 0.47 & 0.2 & 1& 20 & 0.4 &  148 &    - & 4 & - & 3& -& 245 &   - &  48 &  - &  22 &   - &  9 &  - \\
    \hline
    \end{tabular}
    \caption{Col. 1-4: initial number of stars, cluster mass and half-mass radius, primordial binary fraction. Col. 5: number of indipendent realisations. Col. 6-7: initial half-mass relaxation and segregation time. Col. 8: simulated time. Col. 9-10: number of compact object mergers inside the cluster. Col. 11: maximum BH mass during the simulation. Col. 12: maximum BH mass at the end of the simulation. Col. 13-14: number of BHs with a mass $m_{\rm BH}>30\Ms$ or $>40\Ms$ at the last simulation snapshot.}
    \label{tab:t1}
\end{table*}

\subsubsection{Component masses and formation channels}

The population of compact binary mergers in \dragonii consists in 75 BH-BH, 2 NS-BH, and 1 WD-BH. Among BH-BH mergers, 45 involve two BHs below the PPISN maximum mass ($m_{\rm BH} < 40.5\Ms$), 12 involve two mass-gap BHs, and 21 involve one BH below the gap and a mass-gap BH. Six BH-BH mergers involve a primary with a mass $m_{\rm BH,1}=(5.4-7.1)\Ms$ and a companion with mass $m_{\rm BH,2}=(2.55-3.6)\Ms$, i.e. just above the threshold separating NSs and BHs in our models. All these low-mass mergers are in primordial binaries. 
As discussed in paper AS-II, the BHs in the upper-mass gap mostly form in a star-BH accretion event, either by purely dynamical interactions or stellar evolution. We stress that, throughout our models, we assume that a fraction $f_c = 0.5$ of the star mass is accreted onto the BH during an accretion event \citep[for a discussion about the impact of $f_c$ on simulations, see][]{2021MNRAS.501.5257R}.

When GW recoil are ``switched off'' 4 mergers involve a second or third generation BH, i.e. which underwent one or two previous mergers. The inclusion of GW recoil reduces the number of total mergers to 74. For a detailed discussion about the impact of GW recoil, see paper AS-II. 

\begin{figure}
    \centering
    \includegraphics[width=\columnwidth]{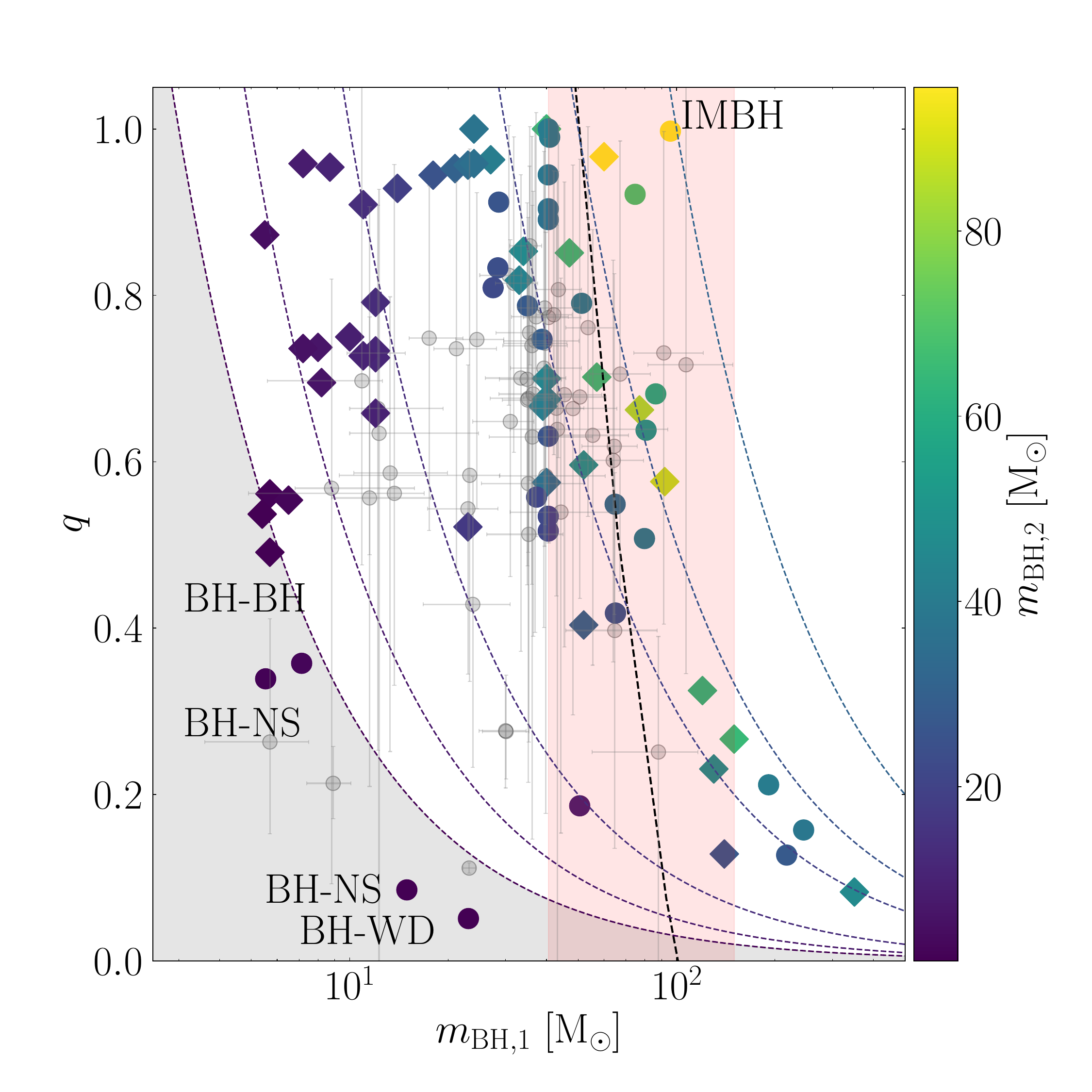}
    \caption{Primary mass (x-axis) and mass-ratio (y-axis) of BH mergers in \dragonii simulations, occurring inside the cluster (points) or after dynamical ejection (diamonds). The colour map identifies the mass of the secondary. The dotted lines corresponds to a companion mass of $m_{\rm BH,2} = 3, ~5, ~10, ~30, ~50, ~100\Ms$. Mergers with a primary mass on the right side of the black dashed line produce an IMBH. The red shaded area roughly identifies the upper-mass gap region. The grey area identify mergers containing a BH and another type of compact object: either a BH in the putative lower mass-gap, a NS or a WD. Shaded grey points represent observed BH mergers from the GWTC-3 catalogue \citep{2021arXiv211103634T}.}
    \label{fig:GWsrc}
\end{figure}

Figure \ref{fig:GWsrc} shows the component masses and mass ratio of \dragonii mergers and of mergers observed during the first three LVK observation campaign, collected in the so-called GWTC-3 catalogue \citep{2021arXiv211103634T,2021arXiv211103606T}. The plot includes all mergers occurring in the cluster or outside the cluter after being ejected via dynamical interactions considering zero GW kicks.

This plot illustrates the wealth of information hid in the Dragon-II star clusters: we find mergers in the upper-mass gap, IMBHs\footnote{In this work we set a mass treshold of $M_{\rm IMBH,min} = 100\Ms$ to discern between BHs and IMBHs.}, repeated mergers, and in a handful cases also BHs merging with either a NS or a WD. 
Interestingly, we find that mergers occurring inside the cluster are characterised by a primary with mass $m_{\rm BH,1} > 30\Ms$ and a companion with a mass in the range $m_{\rm BH,2} = (20-50)\Ms$. Conversely, mergers occurring outside the cluster --- or ejected mergers --- are characterised by a mass-ratio $q>0.6$ and a primary mass typically $m_{\rm BH,1} < 40\Ms$.

The number of mergers occurring inside the cluster (31) is comparable to that of binaries that merge after being ejected from the cluster (47), thus suggesting that in-cluster mergers can made-up the $40\%$ of the total merger population. Among all of them, 27 are from primordial binaries (3 inside, 24 ejected), whilst 51 (28 inside, 23 ejected) are from dynamical binaries. 

\begin{figure}
    \centering
    \includegraphics[width=\columnwidth]{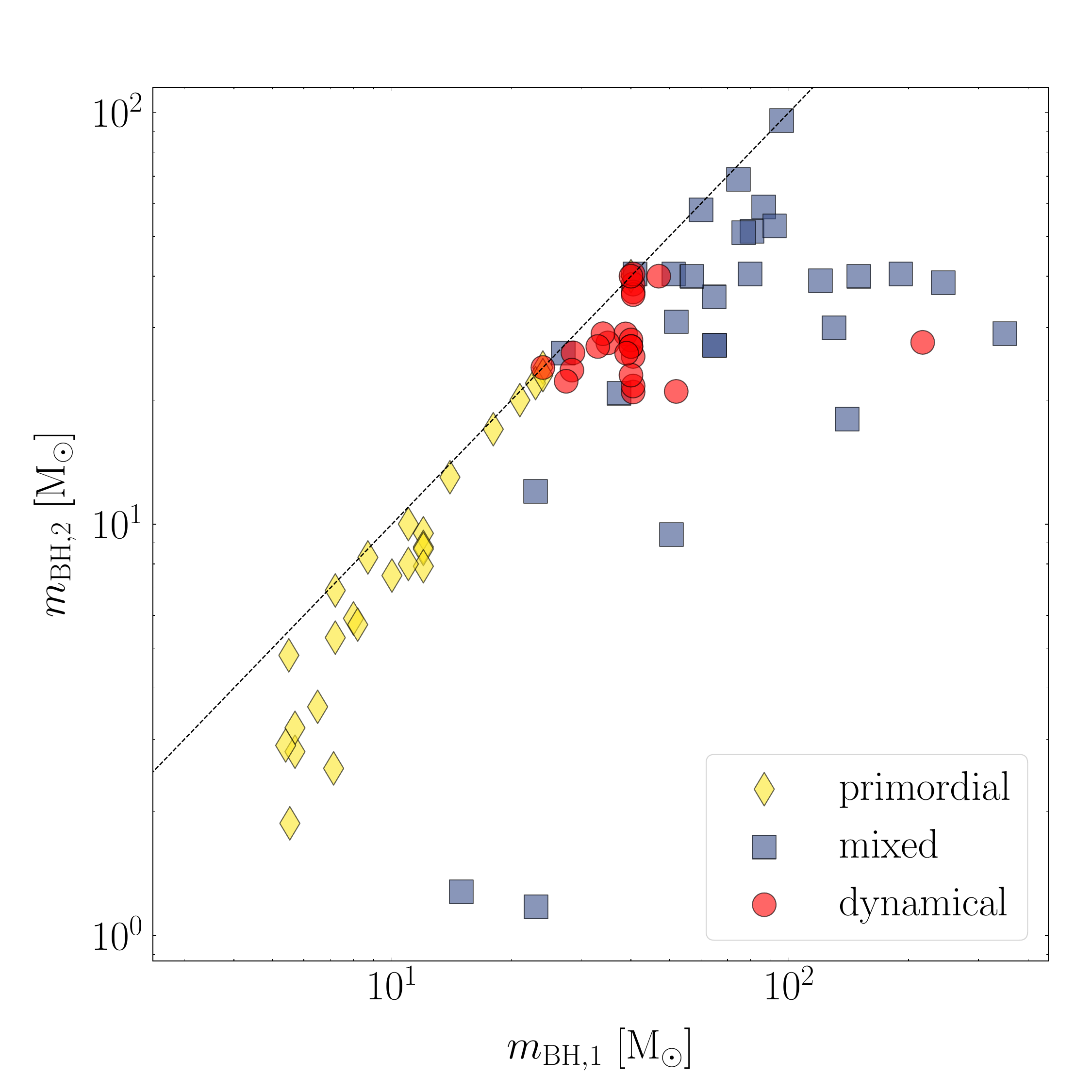}
    \caption{Masses of the primary and companion merging BHs from primordial binaries (yellow diamonds), dynamical binaries (red circles), and binaries with at least one component being a former primordial binary member (blue squares).}
    \label{fig:gwimp}
\end{figure}

Figure \ref{fig:gwimp} shows the primary and companion mass of mergers originated from primordial, dynamical, or mixed binaries, with the latter identifying binary mergers in which at least one component was originally in a primordial binary. The plot exhibits some interesting features: 1) mergers from primordial binaries tend to have nearly equal-mass components, 2) purely dynamical mergers have masses that occupy a tight region of the plane with $m_{\rm BH,1}=(20-50)\Ms$ and $m_{\rm BH,2}=(20-40)\Ms$, 3) mergers with one component previously in a primordial binary are characterised by a heavy primary, $m_{\rm BH, 1} > 40\Ms$, and a heavy companion, $m_{\rm BH,2} > 20\Ms$. A similar trend is observed in recent $N$-body simulations tailored to relatively light star clusters, i.e. with mass $<8,000\Ms$ \citep{2022MNRAS.517.2953T}.

As deeply discussed in paper AS-II, the crucial role of primordial binary dynamics is highlighted by the fact that all the IMBHs in \dragonii clusters but one have an ancestor that was member of a primordial binary, regardless of the IMBH formation scenario. 

Dynamics and binary stellar evolution deeply impact also the properties of stellar-size mergers. For example, ``dynamical'' and ``primordial'' mergers occupy two well separated regions of the primary mass - mass ratio plane.

The vast majority of primordial binary mergers occupy a region delimited by $q > 0.6$ and $m_1 = (5-40)\Ms$, with the mass-ratio weakly increasing at increasing the primary mass: note that for $m_1\lesssim 15\Ms$ mergers have mass-ratio $q=0.6-1$, whilst mergers with a heavier primary have mass ratio $q>0.85$.

Dynamical mergers, instead, form in the right hand-side of Figure \ref{fig:gwimp}, generally at $m_1 > 40.5\Ms$. In this case, the mass ratio decreases with the primary mass as expected from the mass function limit, with companion masses in the range $m_2 = (30-50)\Ms$.

We can identify three relatively well separated regions: low BH masses ($m_{\rm BH,1} <15\Ms$) and widely distributed mass ratio ($q=0.6-1$) dominated by primordial binary mergers, BH masses in the range $m_{\rm BH,1} = (15-40.5)\Ms$ and high mass ratios ($q>0.9$) dominated by primordial binary mergers, and heavy BH primaries ($m_{\rm BH,1}>40.5\Ms$) with relatively massive companions ($m_2=30-50\Ms$) dominated by dynamical mergers.

In \dragonii clusters, most binaries merging outside the cluster originate from primordial binaries and their ejection is typically triggered by the BH natal kick. However, all ejected mergers with component masses $m_{1,2} > 30\Ms$ have a dynamical origin, owing to the adopted stellar evolution recipes. 

We note that, given the limited simulation time, the population of mergers in \dragonii clusters may lack some element that could form later in the cluster life, beyond several relaxation times. These late mergers would unavoidably have a dynamical origin, or at most a "mixed" origin, because all BHs formed in primordial binaries undergo a binary exchange or have been ejected in \dragonii clusters over the simulated time. Moreover, late mergers will likely have smaller masses compared to those shown in Figure \ref{fig:gwimp}. This is mostly due to the BH-burning process, by which the average BH mass decreases over time \citep[see e.g. paper AS-I;][]{2015PhRvL.115e1101R,2017ApJ...836L..26C}. 
As a consequence, some BH mergers forming at late time may have properties similar to the primordial binary mergers shown in Figure \ref{fig:gwimp}.

\begin{figure}
    \centering
    \includegraphics[width=\columnwidth]{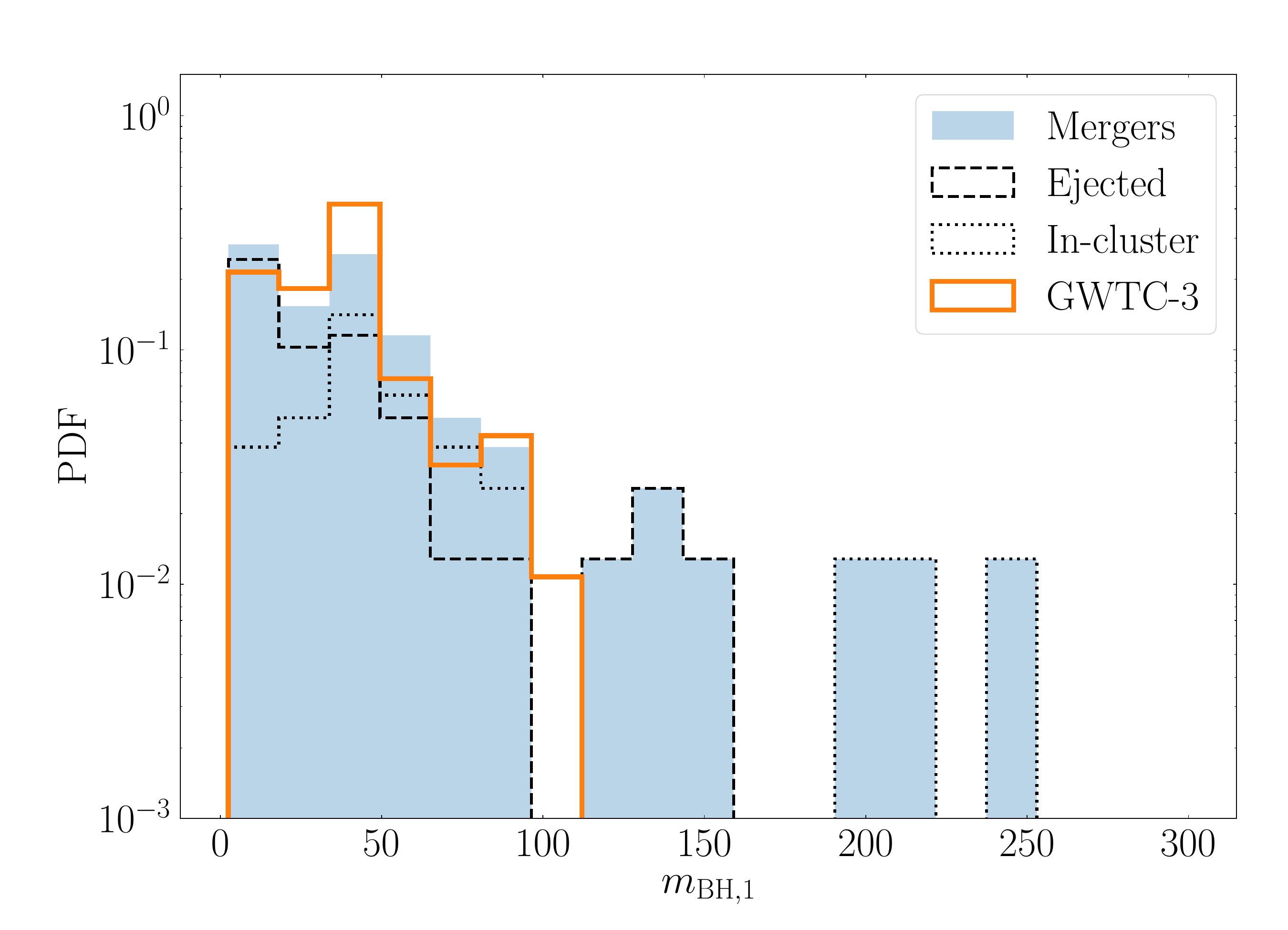}\\
    \includegraphics[width=\columnwidth]{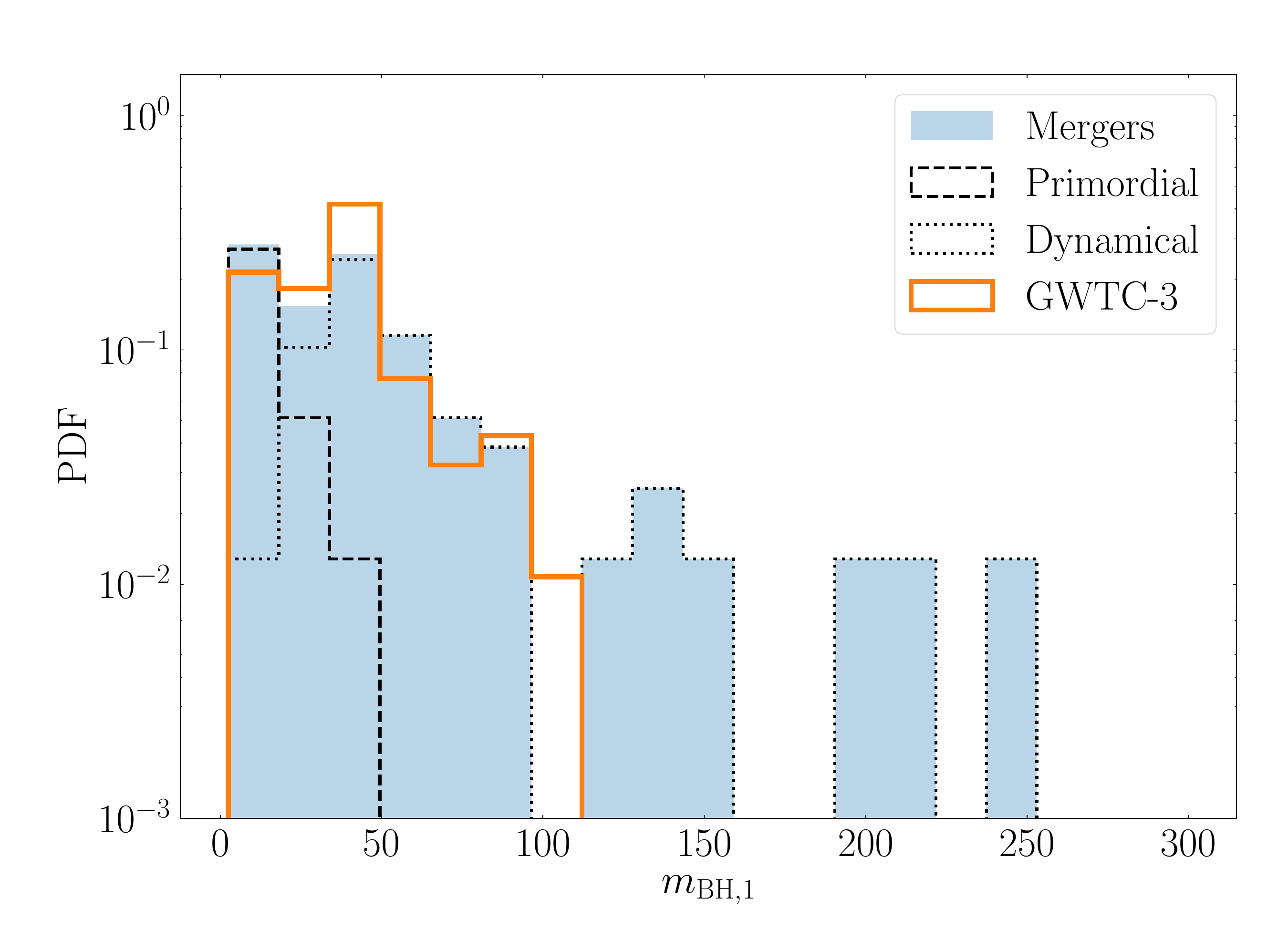}
    \caption{Mass distribution of primary BH mergers from the GWTC-3 catalogue \citep{2021arXiv211103606T} (orange straight line) and from \dragonii simulations (filled light blue steps). The BH mass distribution inferred from LVK data is overlaid to the histograms (black line), with the shaded area ancompassing the $90\%$ credible level. Top panel: Simulated mergers are dissected into those occurring inside the cluster (dotted black line) and after dynamical ejection (dashed black line). Bottom panel: Simulated mergers are dissected into those forming from primordial binaries (dotted black line) or via dynamical interactions (dashed black line).}
    \label{fig:fM1}
\end{figure}

Figure \ref{fig:fM1} shows the mass distribution of the primary BH in \dragonii mergers, dissected into in-cluster/ejected mergers and primordial/dynamical ones. Ejected binaries dominate the $m_{\rm BH,1} \lesssim 20\Ms$ mass range, whilst at larger primary masses their number and distribution is similar to that of in-cluster mergers. Dynamical mergers completely dominate the population of mergers with $m_{\rm BH,1} > 20\Ms$, while primordial mergers dominate the population of lighter mergers. Noteworthy, we see that the primary mass distribution for \dragonii mergers nicely overlap with the sample of mergers in the GWTC-3 catalogue, i.e. the catalogue of BBH mergers detected by the LVK collaboration \citep{2021arXiv211103606T}. However, a thorough comparison between modelled and observed mergers would require to take into account observation biases \citep[see e.g.][]{2017ApJ...851L..25F,2020ApJ...894..133A,2021arXiv210912119A}. For this reason, we also overlay to our data the cosmic BH mass distribution inferred from GW detections. 

Comparing models and observations can be crucial to assess the impact of different formation channels on the population of BH-BH mergers \cite[see e.g.][]{2019MNRAS.482.2991A,2020ApJ...894..133A,2020A&A...635A..97B,2021ApJ...910..152Z,2021arXiv210912119A,2022MNRAS.511.5797M}. Our \dragonii models suggest, for example, that BH mergers developing in star clusters could produce a substantial amount of mergers from primordial binaries. The progenitor binary could, in some cases, suffer the impact of dynamical interactions which may alter their orbital parameters. Nonetheless, in most cases BH mergers from primordial binaries could represent "isolated binary merger impostors", because they have properties typical of merging binaries developing within the isolated formation scenario but form in a dynamical environment. Taking into account the impact of these sources with a sort of mixed formation channel is crucial to correctly quantify the role of different formation channels in determining the shape of the mass distribution of detected merging BHs \citep[see also][]{2021arXiv210912119A}.

Moreover, \dragonii models highlights the role of dynamics in determining the formation of BH mergers with masses inside, and beyond, the mass-gap, supporting and complementing previous works on the topic based either on smaller, or lower-density, $N$-body cluster models and Monte Carlo simulations \citep[e.g.][]{2020ApJ...903...45K,2020MNRAS.497.1043D,2021ApJ...908L..29G,2022MNRAS.512..884R,2022A&A...665A..20B}. 

\subsubsection{Delay times}
\begin{figure}
    \centering
    \includegraphics[width=\columnwidth]{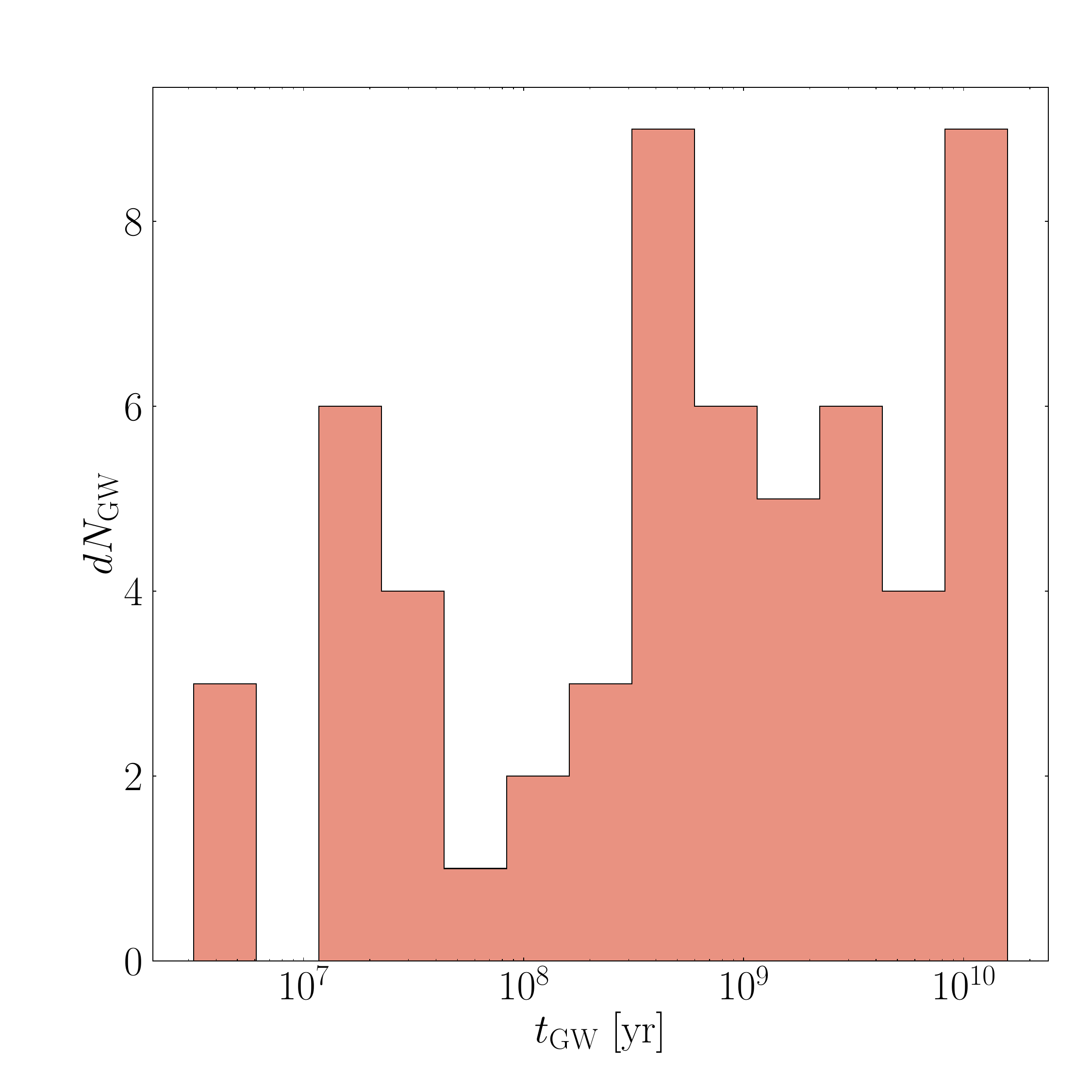}\\
    \includegraphics[width=\columnwidth]{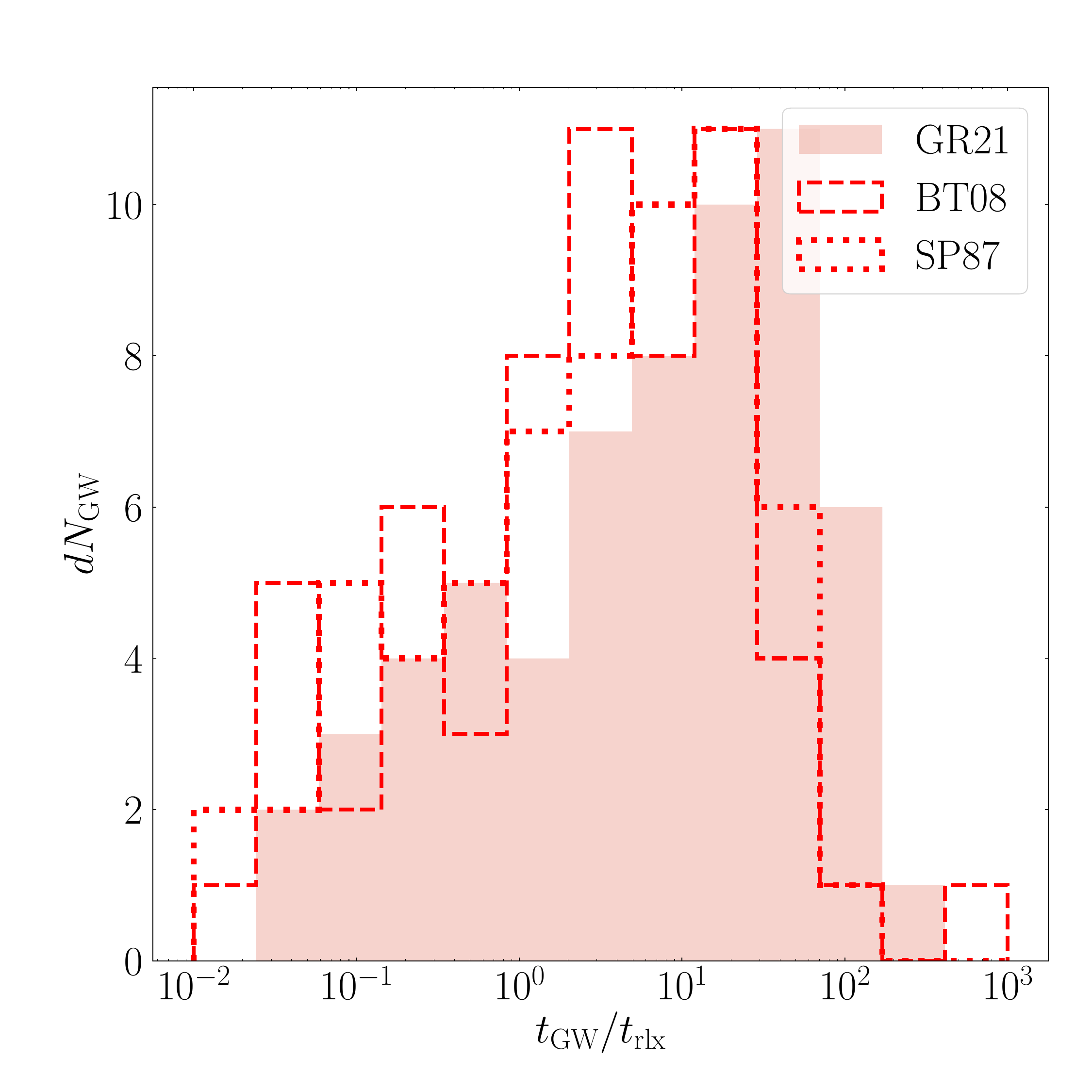}\\
    \caption{Top panel: delay time distribution for \dragonii mergers. Bottom panel: as in top panel, but the time is normalised to the initial cluster relaxation time calculated following \citet{2021MNRAS.507.3312G} (red filled steps), \citet{2008gady.book.....B} (red dashed steps), or \citet{1987degc.book.....S} (red dotted steps).}
    \label{fig:telap}
\end{figure}

The delay time of \dragonii mergers ($t_\gw$), defined as the time elapsed from the beginning of the simulation to the binary merger, is rather peculiar. As show in Figure \ref{fig:telap}, it exhibits three peaks at $t_{\gw} \simeq (0.5 ,~1.5 , ~10)$ Gyr. However, when the delay time is normalised to the initial half-mass relaxation time ($t_{\rm rlx}$) of the cluster, the overall $t_\gw$ nicely distribute around a main peak located at $t_{\rm GW}/t_{\rm rlx} \simeq 8-30$. The exact location of the peak depends on the definition of $t_{\rm rlx}$. For the sake of clarity, in the plot we use three different expressions of $t_{\rm rlx}$ taken from \cite{2021MNRAS.507.3312G} (GR21), \cite{2016ApJ...831..187A} (AR16), or \cite{2021MNRAS.501.5257R} (RN21). 

The three peaks that appear in the $t_\gw$ distribution find a clear explanation looking at the $t_\gw/t_{\rm rlx}$ distribution. In fact, the first peak at $t_{\rm GW} = 500$ Myr corresponds to mergers happening in simulations with $t_{\rm rlx}=50-100$ Myr, whilst the second peak corresponds to mergers occurring in clusters with a longer relaxation time (see Table \ref{tab:t1}). This interesting features suggests, on the one hand, that the delay time depends intrinsically on the cluster initial properties, as they determine the relaxation time, and, on the other hand, that dynamical processes operate in a similar way over a relaxation time regardless of the cluster structure.  The third peak, instead, corresponds to ejected binaries that merge outside the cluster, which are mostly products of primordial binaries ejected via SN explosion during the formation of one of the BHs in the pair.

\subsubsection{Eccentricities}

One intriguing question that arose since the first detection of GWs is whether it is possible to untangle different formation channels in the observed population of BH mergers. Among all the parameters at play, the orbital eccentricity could represent the key to answer this question. Broadly speaking, in fact, most BH mergers forming via binary stellar evolution are expected to feature a negligible eccentricity close to merger, either because the BBH progenitor undergoes common envelope, which shrinks and circularise the orbit, or because the BBH separation is initially so large that GW emission circularise the orbit before the merger. Binaries developing in star clusters, instead, can form with high eccentricity and sufficiently small separation that the merger occurs on a timescale shorter than GW circularisation.

At the lowest-order level, binaries merging in galactic fields, often called isolated binary mergers, are expected to be circular GW sources, whilst at least some of those developing in star clusters and galactic nuclei, named dynamical mergers, can preserve a significant eccentricity (i.e. $e > 0.1$) when entering the typical frequency bands of GW detectors. 

This simplistic division between isolated and dynamical binaries does not take into account several layers of complication. For example, it is well known that star clusters and stellar nurseries may contain a large fraction of binaries, especially among the population of massive stars, where the percentage of paired stars attains values as large as $50-100\%$ \citep{2012Sci...337..444S, 2017ApJS..230...15M}. If primordial binaries evolve on a timescale shorter than the typical timescale of dynamical interactions, star cluster could harbor a sub-population of compact binary mergers with properties pretty similar to those forming in galactic fields, e.g. low eccentricities or peculiar component masses and mass-ratios. 

\begin{figure}
    \centering
    \includegraphics[width=\columnwidth]{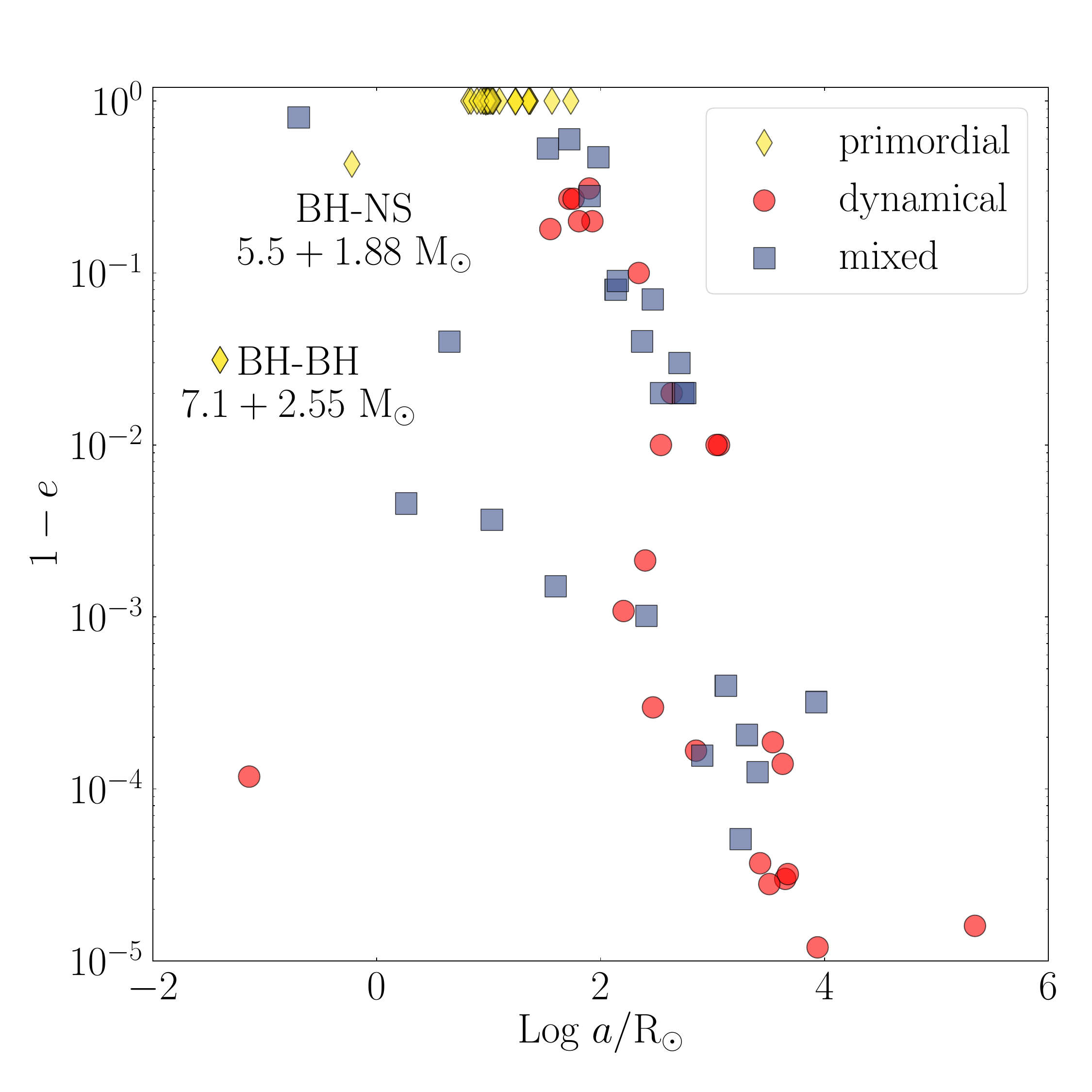}
    \caption{Semimajor axis and eccentricity of compact binary merger at decoupling for primordial (yellow diamonds), dynamical (red points), and mixed (blue squares) mergers. Labels indicate particularly interesting primordial mergers with a significant eccentricity at decoupling.}
    \label{fig:smaecc}
\end{figure}

With up to $33\%$ of stars initially paired, \dragonii simulations offer us the possibility to search for differences between mergers forming entirely via dynamics and those forming from the evolution of primordial binaries. Figure \ref{fig:smaecc} shows the semimajor axis and eccentricity of all BH-BH mergers in \dragonii clusters calculated at the moment of decoupling, i.e. when the GW emission starts dominating over dynamical perturbations. The plot dissects the \dragonii population of BH mergers into those coming from the evolution of primordial binaries, those assembled purely via dynamical interactions, and those involving at least one component that was former member of a primordial binary. The population of dynamical and mixed binaries seem to follow two different sequences, although the low statistics make hard to understand whether they actually exist. The population of nearly circular primordial binaries is evident. These mergers can be considered mimickers of the field merger population, and constitute the $33\%$ of the whole population of \dragonii mergers. Only two of the primordial binaries exhibit a significant eccentricity and a relatively small separation. 

The first is a NS-BH binary, we postpone a discussion about this specific source to the next subsection. 

The second one involves two low-mass BHs, with masses $m_{\rm BH1,2}=(7.1+2.55)\Ms$ and eccentricity $e=0.997$. The progenitor of this merger was a binary that underwent a common envelope phase first, after which the first BH formed, and later undergo Roche lobe overflow, at the end of which also the second BH forms and receives a small kick ($\sim 3$ km$/$s) that triggers the eccentricity increase. 

\begin{figure}
\includegraphics[width=\columnwidth]{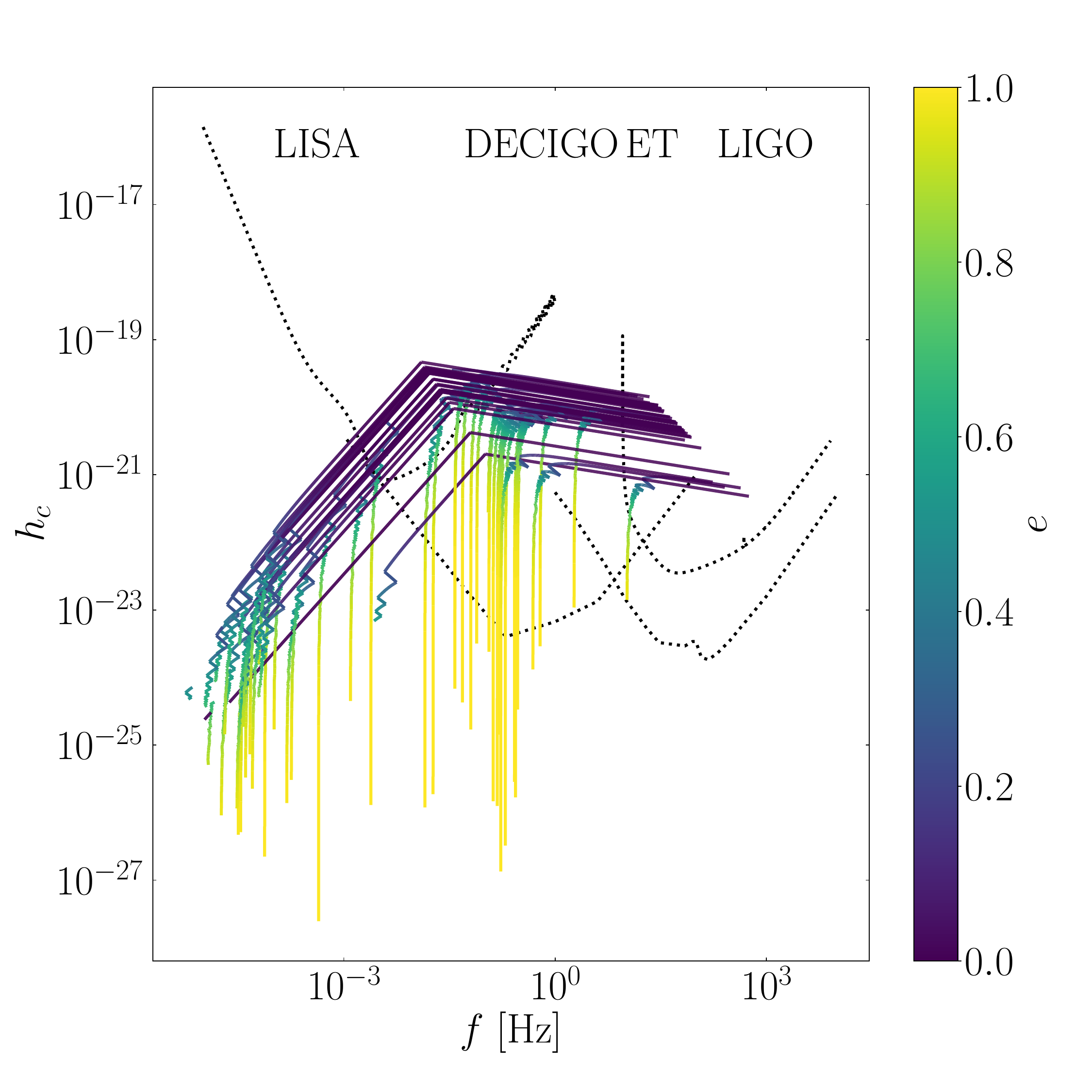}
\caption{Evolution of the binary GW strain as a function of the frequency for all mergers in \dragonii simulations, assuming that the sources are located at a redshift $z=0.05$. The colormap identifies the binary eccentricity along its orbit. Dotted lines represent sensitivity curves for different detectors, from left to right: LISA, DECIGO, ET, and LIGO.}
\label{fig:strain}
\end{figure}

As the binary shrinks and circularises because of GW emission, its frequency will increase. Therefore, a first step to determine whether a binary merger can appear eccentric in the sensitivity band of a specific GW detector requires to compare the binary eccentricity and the corresponding GW frequency. 
We show in Figure \ref{fig:strain} the characteristic strain - frequency evolution for all mergers in our sample, assuming that they are located at a redshift $z = 0.05$, i.e. at a luminosity distance of 230 Mpc. To calculate the GW strain of \dragonii sources we follows the formalism laid out in \cite{1963PhRv..131..435P, 1964PhRv..136.1224P} and the implementation described in \cite{2021A&A...650A.189A} (see Eqs. 30-39). The GW signal from simulated mergers is overlaid to the sensitivity curves of current and future ground-based and space-based detectors like LIGO \citep{2015CQGra..32b4001A}, Einstein Telescope \citep[ET][]{2010CQGra..27s4002P,2023arXiv230315923B}, DECIGO \citep{2011CQGra..28i4011K,2018PTEP.2018g3E01I,2021PTEP.2021eA105K}, and LISA \citep{2013GWN.....6....4A,2022arXiv220306016A}. The plot highlights how the eccentricity drops as the binary sweeps across different frequency bands.

\begin{figure}
    \centering
    \includegraphics[width=\columnwidth]{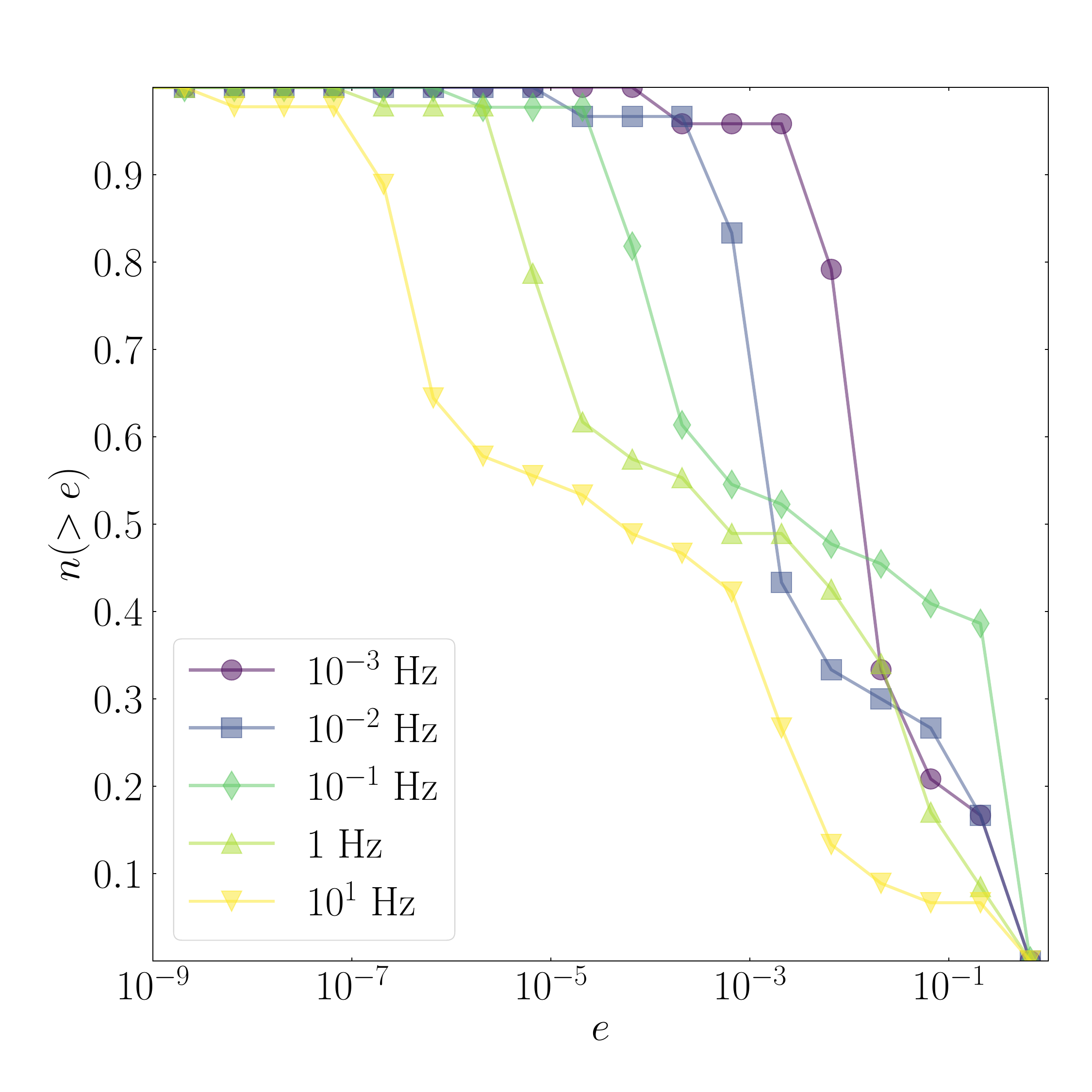}\\
    \includegraphics[width=\columnwidth]{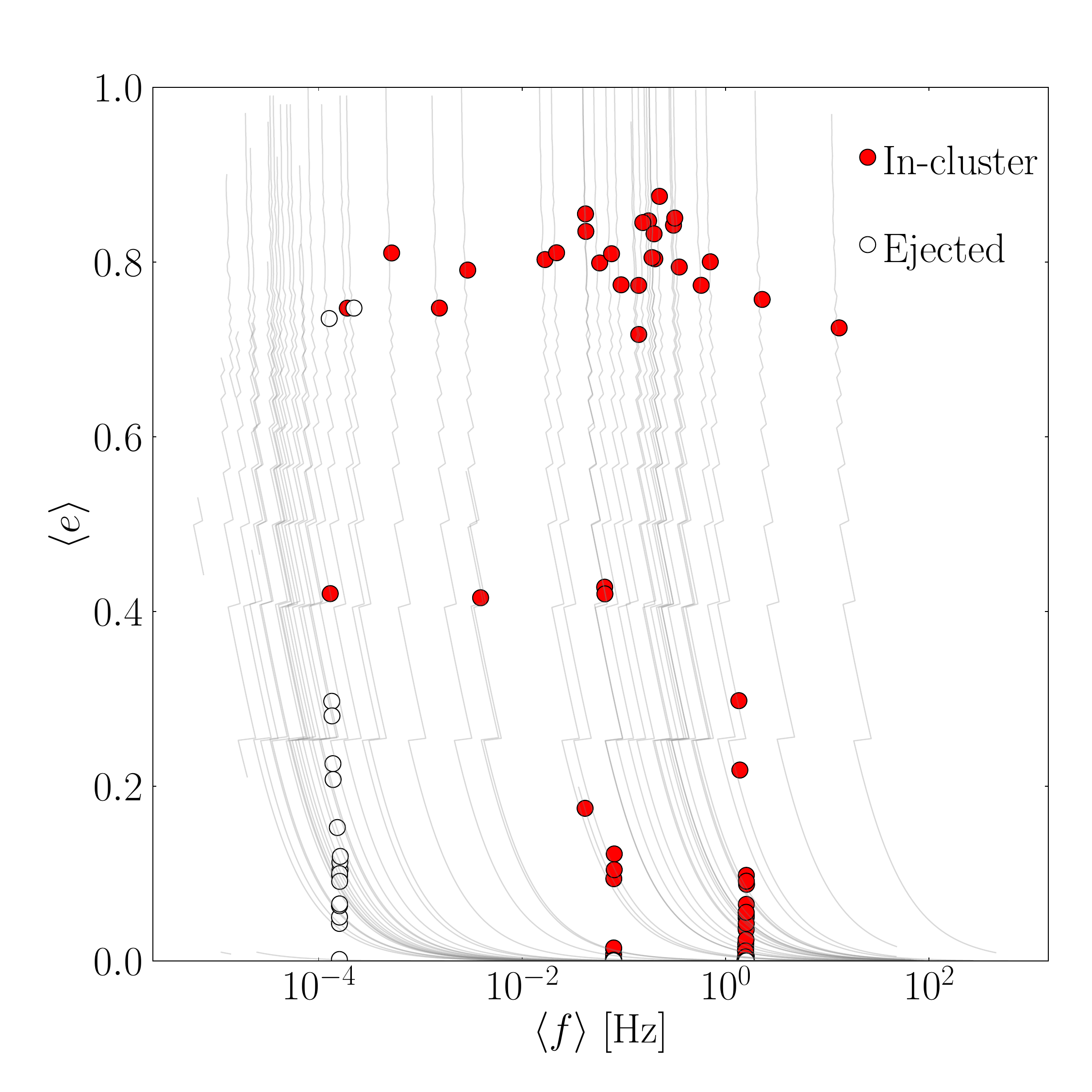}\\
    \caption{Top panel: Fraction of mergers with eccentricity above a given value. Different colors correspond to different frequency bands. Bottom panel: eccentricity evolution as a function of the GW frequency for all mergers in \dragonii models (grey lines) and average value of the eccentricity for mergers occurring inside the cluster (red points) or after ejection (white points).}
    \label{fig:ecce}
\end{figure}

The top panel in Figure \ref{fig:ecce} shows the fraction of mergers with eccentricity above a given threshold calculated when \dragonii mergers sweep through five frequency bands centered in $f_{\rm band} = 10^{-3}-10^{-2}-10^{-1}-1-10$ Hz, i.e. the typical sensitivity bands of space-borne detectors like LISA ($<10^{-2}$ Hz), mid-frequency detectors like DECIGO ($10^{-2}-1$ Hz), and ground-based detectors like LIGO-Virgo-Kagra or the Einstein Telescope ($>1$ Hz). The plot highlights the fact that around $20-40-5\%$ of all mergers appear eccentric, i.e. $e > 0.1$, while moving through the $f = 10^{-3}-10^{-1}-10^{1}$ Hz frequency bands, respectively. Clearly, the detectability of these mergers depend on many parameters, among which the location of the merger and the detector properties. Nonetheless, the plot makes apparent the importance of future deci-Hz detectors in placing constraints on the population of eccentric BBH mergers. Moreover, comparing models with future observations will help to quantify the impact of star cluster dynamics on the cosmic population of merging BHs. 

Noteworthy, the eccentricity carries information about the formation history of the merger. For example, we find that all mergers with an eccentricity $e>0.1$ in both the $0.05-1$ Hz and $1-10$ Hz frequency bands occur inside the cluster. The number of eccentric binaries doubles in the $10^{-2}-1$ Hz frequency band, but these eccentric binaries appear almost circular while reaching the ground-based detector band, explaining why it is more likely to find a \dragonii merging binary with significant eccentricity while sweeping through the deci-Hz band. 

Any binary merger will spend some time in the detector band before merging. In order to characterise the evolution of the eccentricity as the binary inspirals, we calculate the average binary eccentricity weighted with the time to the inspiral, i.e. $\langle e \rangle = \int_0^{t_{\rm merg}} e dt / \int_0^{t_{\rm merg}} dt$. Practically, we measure the binary eccentricity in subsequent time bins from the time of decoupling to the time of merger and weight it with the remaining time to the merger.  This quantity is shown for all mergers in the bottom panel of Figure \ref{fig:ecce}, along with the evolution of the eccentricity as a function of the peak frequency \citep[e.g.][]{2009MNRAS.395.2127O}  
\begin{align}
f_p =& 0.29{\rm Hz} \left(\frac{m_1+m_2}{30\Ms}\right)^{1/2}\left(\frac{a}{50{\rm R}_\odot}\right)^{-3/2} \times \nonumber \\
& \times {\rm ceil}\left[1.15\frac{(1+e)^{1/2}}{(1-e)^{3/2}}\right],
\label{eq:fpeak}
\end{align}
The step-like behaviour of the $e-f_p$ is due to the ceil function in Equation \ref{eq:fpeak}, which returns the nearest integer larger than the function argument. 
The majority of in-cluster mergers clearly show an average eccentricity $\langle e \rangle > 0.7$ across the whole 0.01-100 Hz frequency spectrum, whilst ejected mergers preserve a moderate eccentricity $\langle e \rangle < 0.4$ in the $f<1$ mHz band. This suggests that GW detectors operating in different bands can probe different sub-populations of mergers forming in dense star clusters, with high-frequency detectors being more suited to observe short-lived, highly eccentric mergers occurring inside star clusters, and low-frequency detectors more suited to observe GW sources merging outside their parent cluster.

\subsection{Exotic mergers}
\label{sec:exo}
Despite the relatively small simulation grid, we also find some {\it exotic} mergers: a dynamical WD-BH and 2 NS-BH mergers, one dynamical and one from a primordial binary. The three mergers occur in the most dense simulations in our sample: the WD-BH merger occurs in a simulation with $N=120$k, $R_\ham=0.47$pc, the dynamical BH mergers develop in a simulation with $N=300$k, $R_\ham=0.47$pc, and the one forming from a primordial binary in a simulation with $N=120$k, $R_\ham=0.8$pc. 
This type of mergers are particularly rare in star clusters, because dynamical exchanges favor the replacement of the light component with another BH. Given their rarity, we discuss in the following the details of the formation and evolution of these interesting sources.

\subsubsection{White dwarf - black hole mergers: implications for low-mass X-ray binaries}
\begin{figure}
    \centering
    \includegraphics[width=\columnwidth]{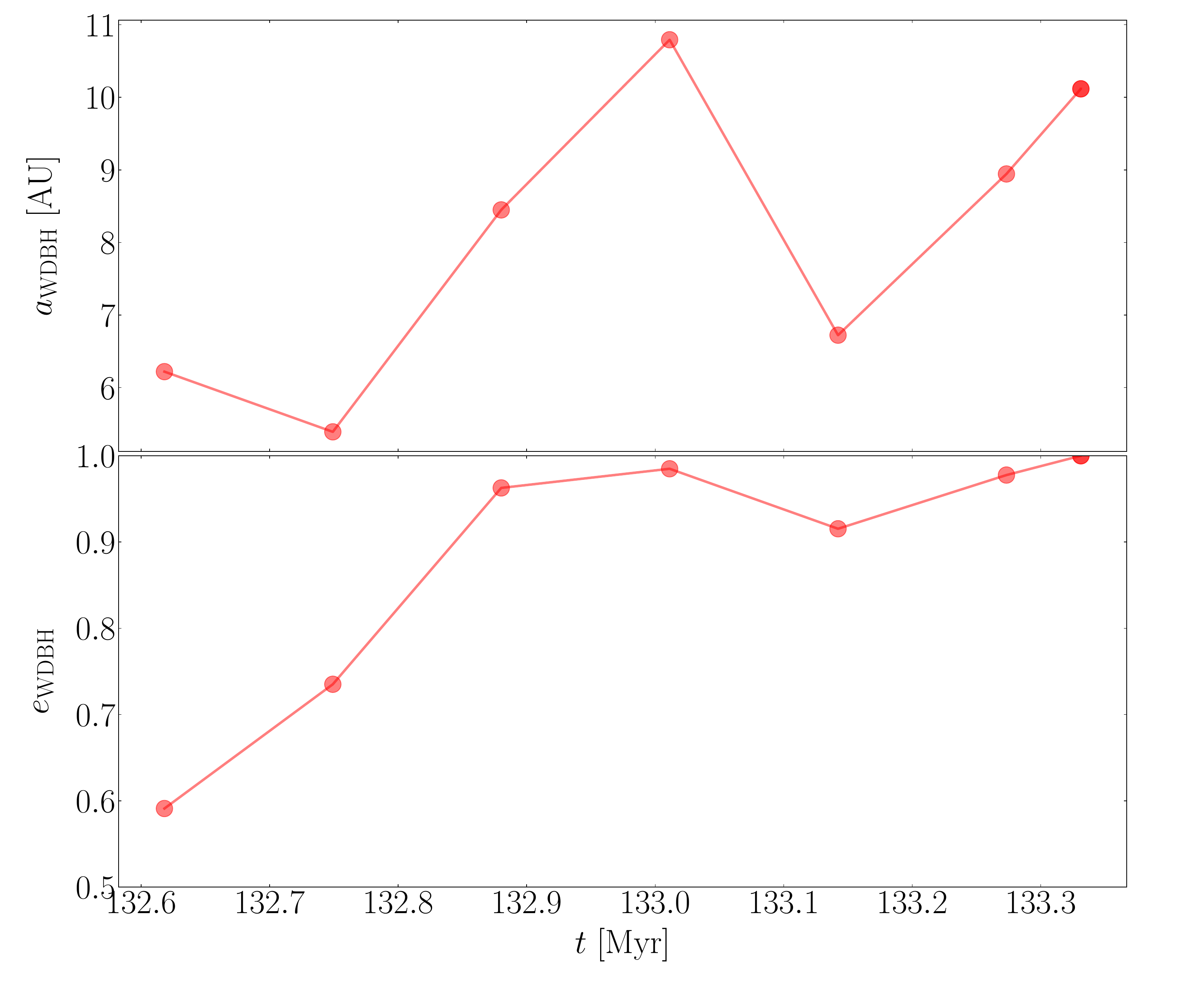}
    \caption{Time evolution of the semimajor axis (top panel) and eccentricity (bottom panel) of a dynamical WD-BH merger in \dragonii clusters. }
    \label{fig:wdbh}
\end{figure}

The WD-BH binary consists in a BH with mass $m_{\rm BH} = 23.1\Ms$ and a carbon-oxygen white dwarf (COWD) with mass $m_{\rm WD} = 1.18\Ms$. Initially, dynamical interactions pair the BH with the WD progenitor, a MS star with mass $m_{\rm MS,pro} = 4.89\Ms$. The two objects undergo common envelope during the late AGB phase of the companion, at the end of which the star turns into a WD, after $\sim 105$ Myr. The resulting WD-BH binary has an "initial" eccentricity of $e = 0.2$ and period of $900$ days. The binary undergoes a series of strong scatterings that cause a rather chaotic variation of the binary semimajor axis and a systematic increase of the eccentricity from $e=0.6$ up to $e=0.99994930$ after 135 Myr, corresponding to $\sim 4$ relaxation times. At this stage, GW emission becomes sufficiently effective to drive binary coalescence. Figure \ref{fig:wdbh} shows the time variation of the WD-BH binary semimajor axis and eccentricity before coalescence. 

The WD Roche lobe is larger than the BH innermost stable circular orbit, hence the WD will likely undergo disruption and start feeding the BH, possibly evolving into a low-mass X-ray binary. In these regards, it is interesting noting the observation of a {\it ultracompact} X-ray binary in the galactic cluster 47 Tuc \citep{2015MNRAS.453.3918M,2017MNRAS.467.2199B}, likely comprised of a COWD and a BH \citep{2017MNRAS.467.2199B}, with the BH being probably heavier than $m_{\rm BH} > 9\Ms$ \citep{2017ApJ...851L...4C}. 
Our model confirms the possibility to form such type of low-mass X-ray binary via interactions of stars and BHs in a dense cluster, even in a relatively short time ($t < 200$ Myr).
Ultimately, the binary shrinkage driven by GW emission will disrupt the WD and the mass falling onto the BH could possibly power jets that can give rise to transients with peak energy $10^{47}-10^{50}$ erg$^{-1}$ and duration of a few minutes \citep{2019MNRAS.488..259F}, giving rise to a tidal disruption event (TDE). 
Despite this source is the only one undergoing coalescence, we find a total of 50 WD-BH binaries by the end of the simulation in all \dragonii clusters. None of them have orbits such to trigger a TDE within a Hubble time, unless a strong interaction with some cluster member pushes the WD on an extremely eccentric orbit. Pushing the orbit to at least $e > 0.9999$($0.99999$) would lead to 1(26) further WD-BH mergers. 
Note that the eccentricity value required to trigger a WD TDE may seem extreme, but it is comparable to the eccentricity achieved by the WD-BH merger, hence testifying that it is possible to reach such extreme eccentricity values in \dragonii clusters.

\subsubsection{Neutron star - black hole mergers: implications for multimessenger astronomy}

Concerning NS-BH binaries, we find two mergers, one of dynamical origin and the other forming from the evolution of a primordial binary. 

The dynamical NS-BH has a NS with mass $m_{\rm NS}=1.28\Ms$ and a BH with mass $m_{\rm BH} = 14.96\Ms$.
The BH, whose progenitor had a mass $m_{\rm MS} = 26.7\Ms$, undergoes a series of chaotic interactions with a primordial binary containing the NS and its companion, which eventually leads to the merger. When the binary decouples from the cluster dynamics, it has a semimajor axis of $a = 0.33$ AU and an eccentricity $e = 0.99778817$, corresponding to a GW peak frequency $f_{\rm GW} = 0.01$ Hz.

After decoupling, the binary evolution is completely dominated by GW emission and the variation of its orbital parameters can be described, at first order, via the \cite{1964PhRv..136.1224P} formalism. We find that as the binary sweeps through the $0.01-0.5-1-10$ Hz GW frequency band the NS-BH merger has a residual eccentricity of $e_{\rm NSBH} = 0.99779-0.9974-0.21-0.02$, thus in principle measurable with future GW detectors, especially with those operating in the deci-Hz frequency band. 
The chirp mass of this merger, $\mathcal{M}_{\rm chirp} = 3.4\Ms$, is typical of dynamically assembled NS-BH mergers \citep{2020CmPhy...3...43A,2020MNRAS.497.1563R,2020ApJ...888L..10Y,2021ApJ...908L..38A}, but hard to produce with isolated binary evolution \citep{2018MNRAS.480.2011G,2020ApJ...899L...1Z}, although this strongly depends on the adopted stellar evolution scheme \citep{2021MNRAS.508.5028B}. 

The primordial NS-BH binary merger, instead, forms from a primordial binary with initial mass components $m_{1,2} = (26.3 + 18.7) \Ms$ and evolves through a common envelope phase initiated by the primary, which eventually forms the BH. Shortly after, the binary undergoes a second common envelope phase and eventually the companion evolves into a NS. Eventually, the merging binary consists of a BH with mass $m_{\rm BH} = 5.6\Ms$ and a NS mass $m_{\rm NS}=1.88\Ms$. Note that these properties, are intriguingly similar to GW200115, a GW source detected by the LVK during the O3 observation campaign, which was characterised by a BH with $m_{\rm BH}= 5.7^{+1.8}_{-2.1}\Ms$ and a NS with $m_{\rm NS} = 1.5^{+0.7}_{-0.3}\Ms$. 
When the NS forms, the common envelope has shrunk the binary from $2.5$ R$_\odot$ to $a = 0.6$ R$_\odot$, whilst the natal kick imparted at formation onto the NS causes an enhancement of the eccentricity from nearly zero to $e= 0.57$. The new orbital parameter as such that GW emission dominates over dynamics and the binary coalesces in $\sim 7\times 10^4$ yr. At decoupling, the binary peak frequency is $f_{\gw} \sim 2$ mHz, right in the middle of LISA sensitivity band. 

The development of a NS-BH binary merger from a primordial binary in a dense star cluster highlights the impact of primordial binaries in contributing to the population of mergers with properties similar to those forming in isolation, making quite hard untangling their actual origin. 

Merging NS-BH binaries are thought to be possible progenitors of several electromagnetic (EM) transients, like short Gamma Ray Bursts (sGRBs)  \citep[e.g.][]{2007NJPh....9...17L} and kilonovae \citep[e.g.][]{2012ApJ...746...48M}. A basic condition for the possible development of an EM transient is that part of the NS material remains bound to the BH, forming a disk. The fraction of NS mass in the disk depends on several quantities, among which the BH-to-NS mass ratio $m_{\rm BH}/m_{\rm NS}$, the BH spin $\chi$, and the NS compactness $C\equiv Gm_{\rm NS}/c^2 R_{\rm NS}$ \citep{2012PhRvD..86l4007F}. As numerical simulations have shown, in general the larger the $m_{\rm BH}/m_{\rm NS}$ the larger the minimum spin required for the NS material to form a disk around the BH, and the larger the spin the larger the amount of matter bound to the BH \citep{2012PhRvD..86l4007F}. Depending on the orbital parameters, the BH tidal field can tear apart the NS before it enters the BH event horizon, provided that the NS tidal radius
\begin{equation}
    R_{\rm tid} = R_{\rm NS} \left(\frac{3m_{\rm BH}}{m_{\rm NS}}\right)^{1/3}, 
\end{equation}
exceeds the BH innermost stable circular orbit (ISCO), which for a spinning BH can be expressed as \citep{1972ApJ...178..347B}
\begin{equation}
R_{\rm ISCO} = \frac{Gm_{\rm BH}}{c^2} \left[3 + Z_2 - {\rm sign}(\chi) \left[(3 - Z_1)(3 + Z_1 + 2Z_2 )\right]^{1/2}\right],
\end{equation}
where $Z_{1,2}$ are functions of the BH adimensionless spin parameter $\chi$. Whilst the condition $R_{\rm tid} / R_{\rm ISCO} < 1$ implies that the merger has no EM emission, the opposite does not ensure the EM counterpart detectability, as it depends  on the geometry of the merger with respect to the observer and other possible observation biases.

In \dragonii clusters, the dynamical NS-BH merger is characterised by $m_{\rm BH}/m_{\rm NS} = 11.7$ and compactness $C = 0.19$ (assuming a NS radius of $10$ km). As shown in Figures 6-8 of \cite{2012PhRvD..86l4007F}, the minimum BH spin required for an accretion disk to form with a mass $10\%$ of the NS mass around such type of binary is $\chi_{\rm BH} > 0.98$. The BH formed in this binary did not undergo any major interaction with stellar companions that could spin-up it \citep{2019ApJ...870L..18Q, 2020A&A...635A..97B, 2020A&A...636A.104B}. Hence, it is possible that the BH formed with low-spin, according to the \cite{2019ApJ...881L...1F} model, hampering the formation of a massive accretion disk around the BH and minimizing the probability for a EM counterpart to develop. 
The isolated NS-BH merger, instead, is characterised by $m_{\rm BH}/m_{\rm NS} = 2.98$ and $C = 0.27$. Even in this case, the spin required for an accretion disk to form is $\chi_{\rm BH} > 0.9$. The BH in this binary undergoes a RLO phase, which could, in principle, spin-up the BH up to extremal values \citep[e.g.][]{2015ApJ...800...17F}, although this strongly depends on the stellar evolution recipes and the binary properties \citep{2019ApJ...870L..18Q, 2020A&A...635A..97B, 2020A&A...636A.104B}. 

The development of just 2 NS-BH mergers highlights how rare are these type of objects on the one hand, and make any statistical analysis poor, on the other hand. Nonetheless, the fact that the NS-BH mergers developed in \dragonii clusters seem to be unlikely to feature an EM counterpart supports the idea that most NS-BH mergers proceed unseen in star clusters \citep{2020CmPhy...3...43A}. For comparison, note that for isolated binaries typically $m_{\rm BH}\sim 12\Ms$ and $m_{\rm NS} = 1.6\Ms$ \citep{2021MNRAS.508.5028B}, which implies a minimum BH spin of $\chi_{\rm BH} \gtrsim 0.8$ to permit the formation of a fairly massive (mass > $0.1m_{\rm NS}$) disk around the BH \citep{2012ApJ...749...91F}.

\section{Discussion}
\label{sec:disc}

\subsection{The impact of natal spins on the properties of stellar black hole mergers}
\label{sec:spin}

Spin amplitude and mutual orientation at merger represent two possible quantities that can help discerning whether a BBH merger results from isolated stellar binary evolution or stellar dynamics \citep[e.g.][]{2019MNRAS.482.2991A,2020ApJ...894..133A,2021ApJ...910..152Z,2021arXiv210912119A,2022MNRAS.511.5797M,2022A&A...665A..20B,2023arXiv230210851B}. 

In order to explore the impact of different spin prescriptions on \dragonii mergers, we devise two models.
The first model (hereafter STEV) assumes that the spin is intrinsically related to the BH evolutionary pathways. For BHs formed from single stellar evolution, we assume a negligible spin ($\chi_{\rm BH} = 0.01$) owing to efficient angular momentum transport triggered by the Tayler-Spruit dynamo \citep{2002A&A...381..923S,2019ApJ...881L...1F}. For upper-mass gap BHs formed from massive binary evolution we assume that final spins spans the $\chi_{\rm BH} = 0.8-1$ range \citep{2018A&A...616A..28Q,2020A&A...635A..97B,2020A&A...636A.104B,2020ApJ...892...13S}. For BHs in primordial binaries, instead, we assign to one BH a spin value of $\chi_{\rm BH}=0.01$ and to the other $\chi_{\rm BH} = 0.1-1$ \citep{2018A&A...616A..28Q, 2020A&A...635A..97B}. 

The second model (GAUS model) assumes, instead, that the spin distribution follows a Gaussian distribution with mean $\bar{\chi}_{\rm BH} = 0.5$ and dispersion $\sigma_\chi = 0.2$, regardless the BH past evolution, a case possibly supported by the population of observed BH-BH mergers \citep{2021arXiv211103634T}. 

In our analysis, we assume that the spin vectors in dynamical mergers are isotropically distributed, whilst for primordial mergers we proceeds as follows. We define an ad-hoc distribution function for the cosine of the angle between the spin of the $i$-th binary component and the binary angular momentum, $\theta_i$, such that \citep{2021arXiv210912119A}
\begin{equation}
    P(\cos\theta) = [(\cos \theta + 1)/2]^{n_\theta+1}. 
\end{equation}
We set $n_\theta = 8$, which implies that binaries have a $20(55)\%$ probability to have $\theta_{1,2}$ that differ by less than $5(20)\%$. Note that $n_\theta = 0$ implies the isotropic distribution whilst $n_\theta \gg 1$ implies fully aligned spins, i.e. $\theta_1 = \theta_2$.

\begin{figure*}
    \centering
    \includegraphics[width=0.9\columnwidth]{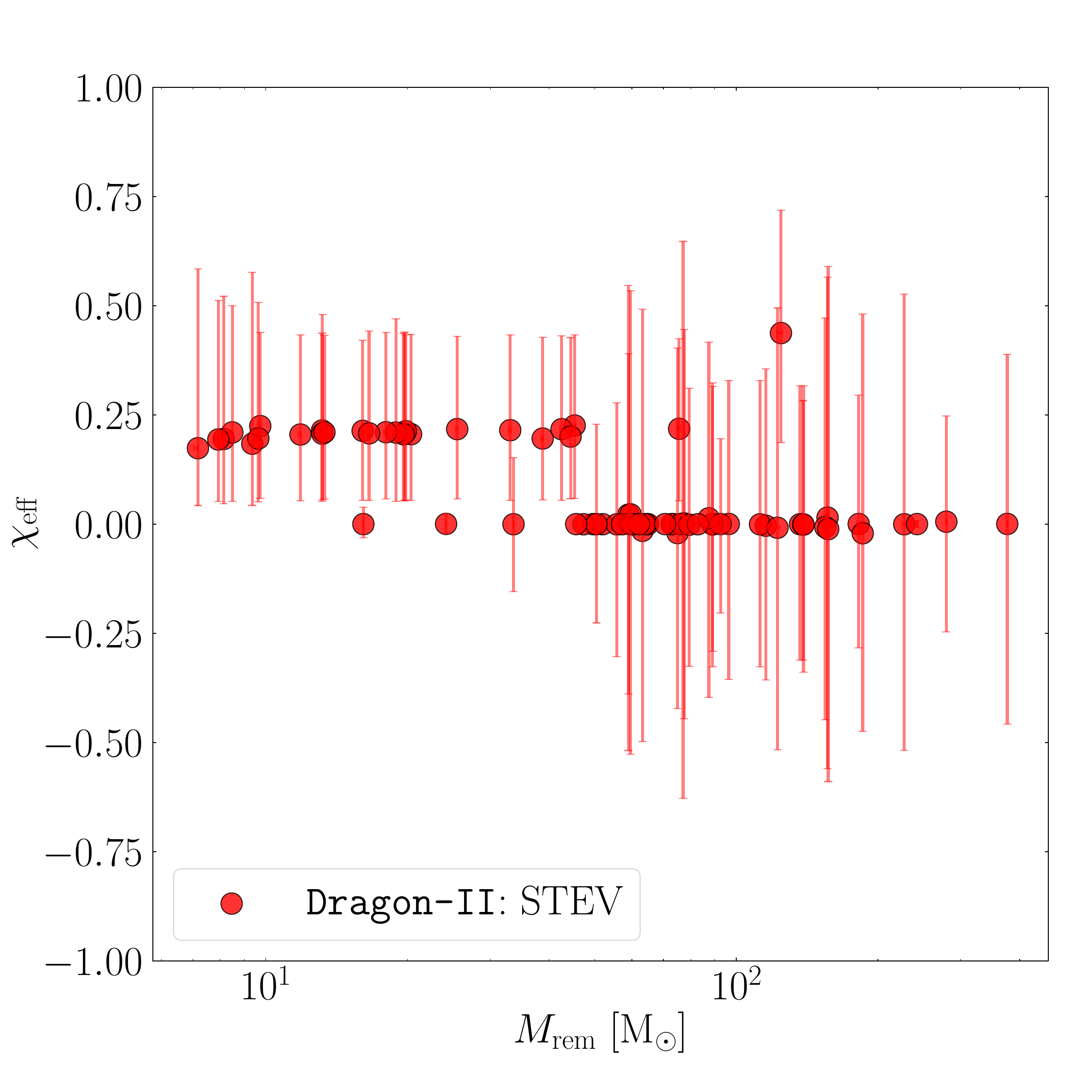}
    \includegraphics[width=0.9\columnwidth]{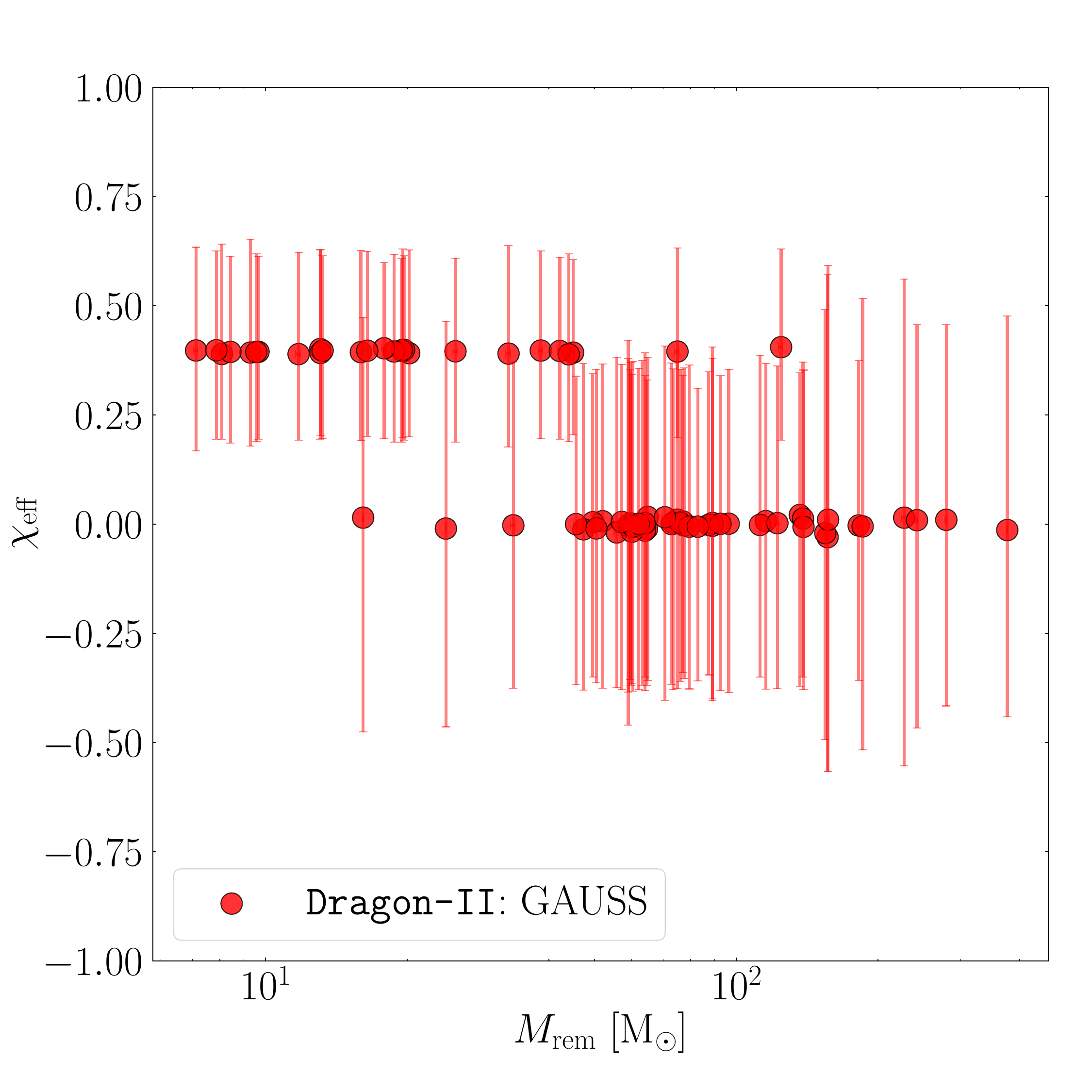}\\
    \includegraphics[width=0.9\columnwidth]{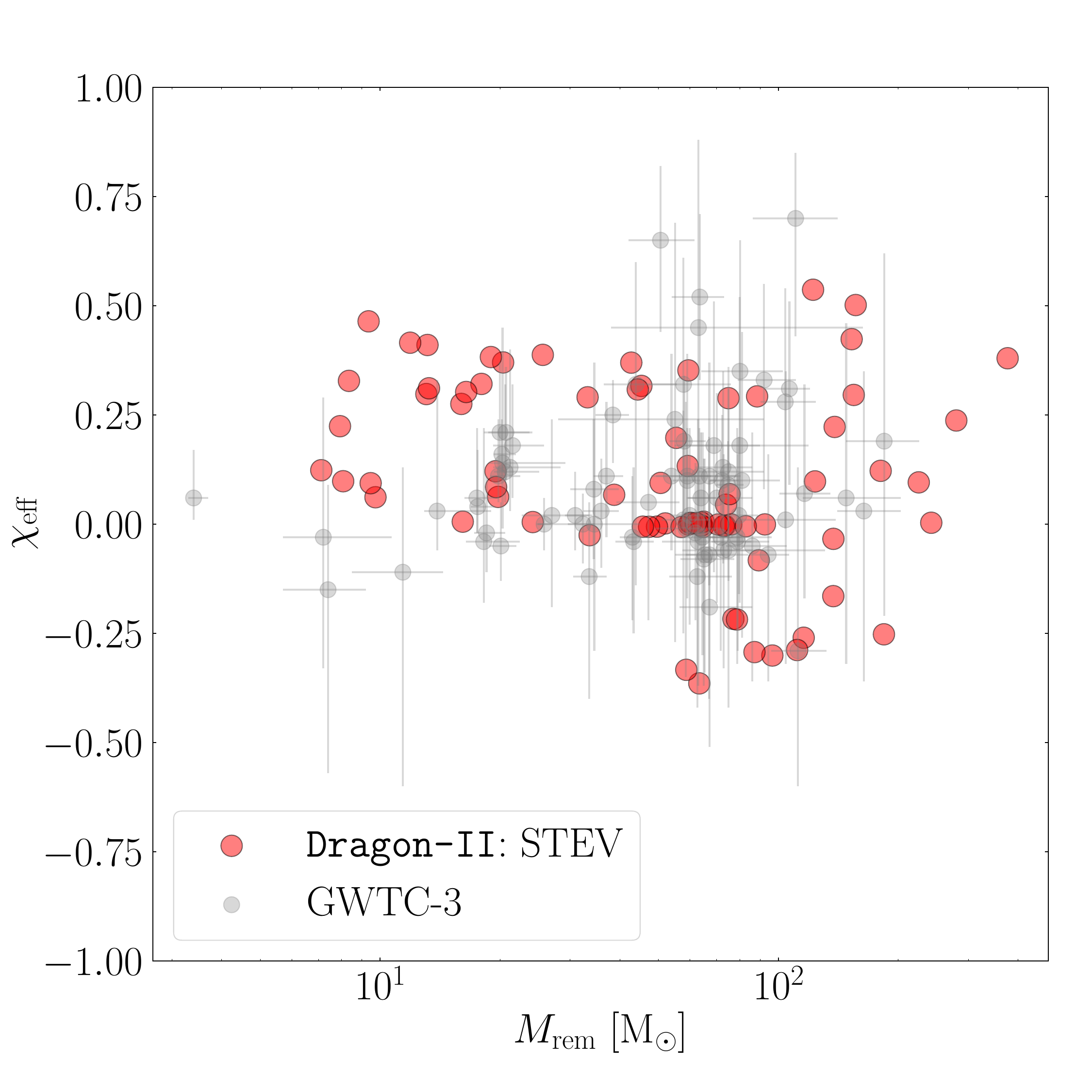}
    \includegraphics[width=0.9\columnwidth]{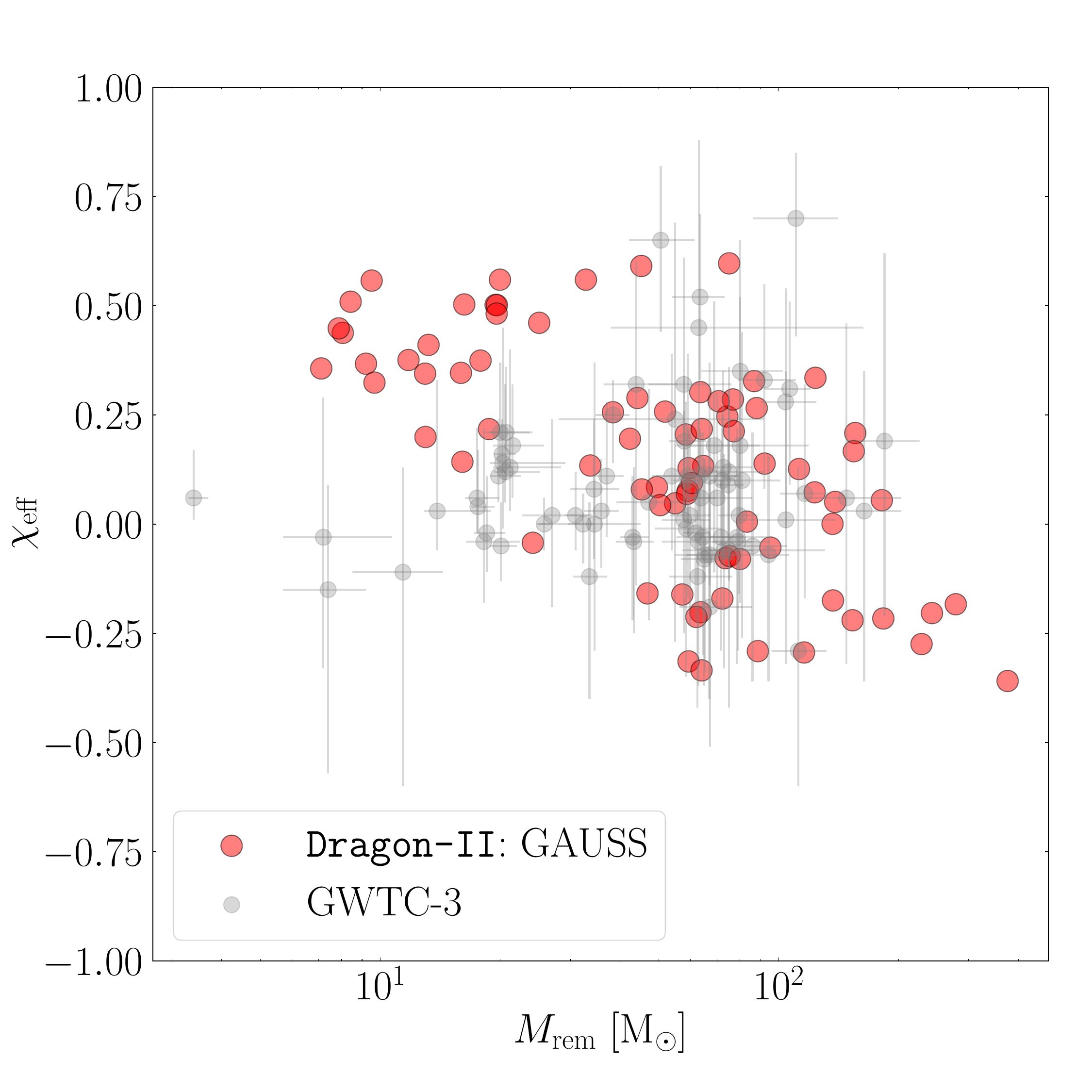}
    \caption{Top panels: median and 95th percentile value of remnant mass and effective spin parameters of \dragonii mergers assuming a prescription for BH natal spin based on stellar evolution models (STEV, left panel) or a Gaussian distribution (GAUSS, right panel). Bottom panel: effective spin parameter and remnant mass for one realisation of \dragonii mergers (red points) and the observed population of LVK mergers in the GWTC-3 catalog.}
    \label{fig:xeff}
\end{figure*}

For each BBH merger in our sample we select 1,000 values of the spin and spin directions depending on the aforementioned assumptions, in order to assess statistically the properties of \dragonii mergers. The top panels in Figure \ref{fig:xeff} show the median value and 95th percentile of the effective spin parameter and remnant BH mass for all BBH mergers in \dragonii models. As, expected, we can clearly see a difference between primordial binaries, which have mildly aligned spins and thus $\chi_{\rm eff}>0$, and dynamical binaries, for which $\chi_{\rm eff} \sim 0$. The plots suggest that the STEV model, based on stellar evolution models, leads primordial binaries to have a $\chi_{\rm eff}$ smaller, on average, compared to the GAUSS model. The bottom panels of Figure \ref{fig:xeff} overlay to a single realisation of the simulated data the observed mergers from GWTC-3, for comparison's sake.

Noteworthy, the assumption that BHs form with a negligible spin unless matter accretion processes are at play (STEV model) leads to a sub-population of mergers with $\chi_{\rm eff} \sim 0$ and $m_{\rm bin} = (40-100)\Ms$, a feature that disappears when a global Gaussian spin distribution is adopted (GAUS model) as shown in Figure \ref{fig:xeff}. 

If BH spins do not strongly depend on stellar evolution processes, but rather are well described by a general distribution, like a Gaussian, we can identify two populations in the plot, one with clearly positive $\chi_{\rm eff}$ values and $m_{\rm BH} < 40\Ms$, and one widely distributed around zero $\chi_{\rm eff}$ involving massive BHs, $m_{\rm BH} > 40\Ms$.

\begin{figure}
    \centering
    \includegraphics[width=\columnwidth]{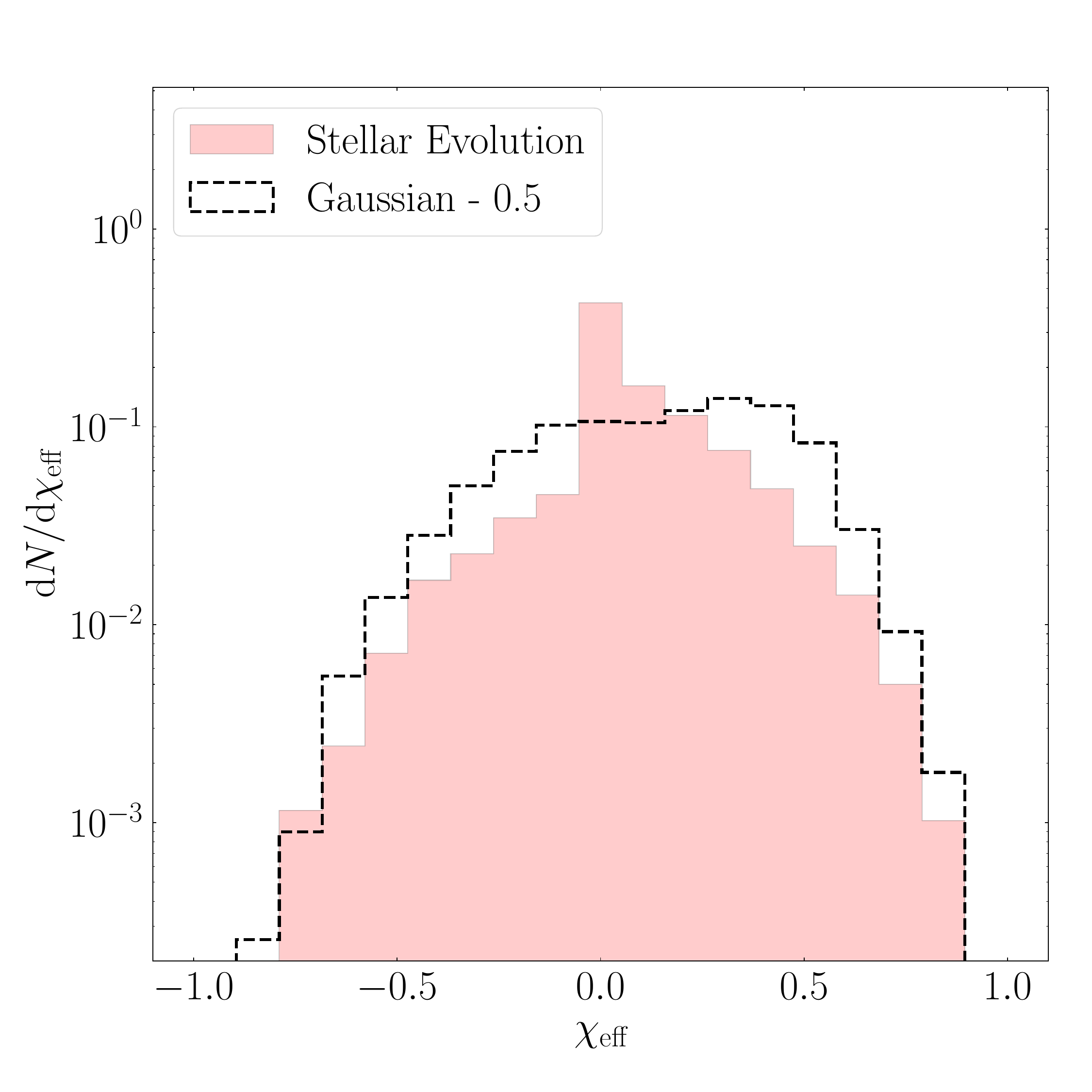}
    \caption{Effective spin distribution for a population of mergers based on \dragonii models assuming model STEV (filled red steps) or GAUS (dashed line).}
    \label{fig:augm}
\end{figure}

In order to improve the poor statistics, we proceed as follows: from the list of \dragonii mergers we create an oversampled catalogue by repeating the spin assignment 100 times and, at each time, selecting a new "mock" BH mass in the range $2.5-40.5\Ms$ if the \dragonii BH merger mass is below the upper-mass gap, and in the range $40.5-100\Ms$ otherwise. This way, each real \dragonii merger will have 100 counterparts with BHs of the same class (upper mass-gap or not, merger in primordial or dynamical binary), but enabling to build-up a catalogue sufficiently rich to analyse the overall $\chi_{\rm eff}$ distribution. Figure \ref{fig:augm} shows the distribution of $\chi_{\rm eff}$ for the augmented sample in STEV and GAUS models.  
We see that the STEV model follows a narrower distribution compared to the GAUS model, and exhibits a clear peak around zero owing to the population of BHs formed from single stars \citep[see also Figure 9 in][]{2021arXiv210912119A}. 

\subsection{Compact binary merger rates}
\label{sec:mergerate}

\subsubsection{Merger efficiency}
As described in the previous section, we have simulated a total \dragonii mass of $M_{\rm sim} = 3.65\times 10^6\Ms$ and find in total 78 mergers when GW recoil is not accounted for, and 74 otherwise. Therefore, the resulting BH merger efficiency, defined as the ratio between the number of mergers and the total simulated mass \citep{2014MNRAS.441.3703Z}, is given by 
\begin{equation}
    \eta_{\rm GW} = \frac{N_{\rm GW}}{M_{\rm sim}} \simeq (2.0-2.1)\times10^{-5} \Ms^{-1},    
\end{equation}
similar to what inferred for young and open clusters with a similar metallicity \citep[e.g.][]{2020MNRAS.498..495D, 2020ApJ...898..152S, 2021MNRAS.507.3612R}. Note that given the limited simulation time our estimate could represent a lower limit to the total merger efficiency in massive young and intermediate-age clusters. Nonetheless, we note that as the cluster loses mass and expands, the binary formation rate and binary--single interaction rate will sharply decrease until the point in which it will be unlikely for tight binaries to form and merge within a Hubble time.

Interestingly, at fixed value of the half-mass radius, the merger efficiency changes sensibly with the initial binary fraction, being
\begin{equation}
    \eta_{\rm GW, fb} = 
    \begin{cases}
       2.3 \times 10^{-5} \Ms^{-1}  & f_b = 0.20, \\
       1.2 \times 10^{-5} \Ms^{-1}  & f_b = 0.05. \\
    \end{cases}
\end{equation}
This highlights the role of primordial binaries in determining the formation of merging compact objects. For comparison, note that the merger efficiency derived in \cite{2021MNRAS.507.3612R} is based on star cluster models containing $\sim 40\%$ of stars in primordial binaries.

\begin{figure}
    \centering
    \includegraphics[width=\columnwidth]{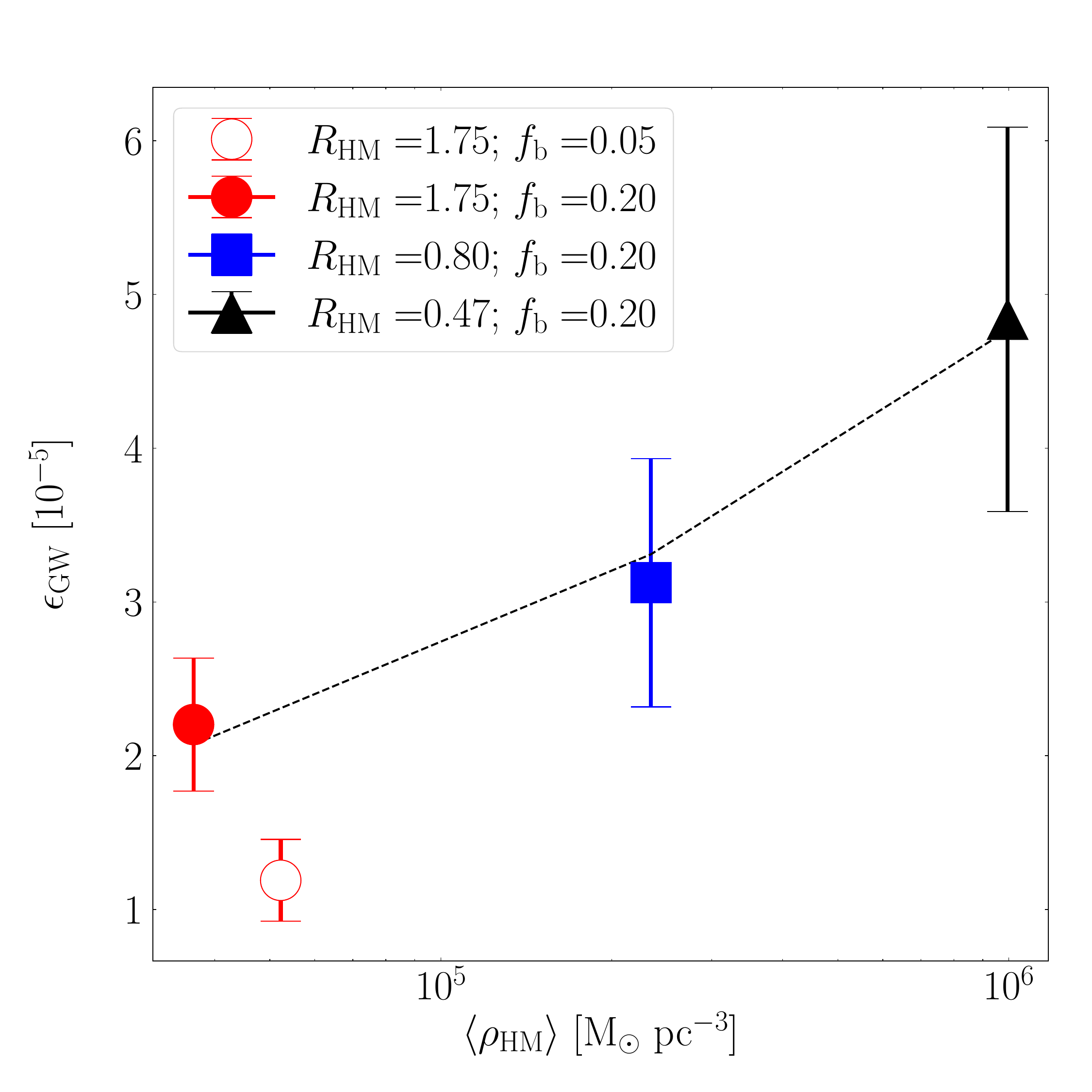}
    \caption{Merger efficiency as a function of the average cluster density. Different symbols correspond to different values of the cluster half-mass radius. Open circles mark simulations with the lowest binary fraction. }
    \label{fig:GWE}
\end{figure}

To further explore the impact of cluster properties on the merger efficiency, we show in Figure \ref{fig:GWE} the average merger efficiency per cluster, $\epsilon_\gw(R_\ham)$, as a function of the average cluster density $\langle \rho_{\rm sim} \rangle$, using the following definitions
\begin{eqnarray}
    \epsilon_\gw(R_\ham) &=& \frac{N_\gw}{(M_{\rm sim}/N_{\rm sim})},\\
    \langle \rho_{\rm sim} \rangle &=& \frac{M_{\rm sim}}{N_{\rm sim}R_\ham^3} 
\end{eqnarray}
where $M_{\rm sim}$ is the total simulated mass and $N_{\rm sim}$ is the number of simulations performed for a given value of the half-mass radius, $R_\ham$. 
At fixed value of the binary fraction, this relation is well described by a power-law in the form $\epsilon_\gw = a (\langle \rho_\ham \rangle / 1{\rm M}_\odot{\rm pc}^{-3}) ^ b$, with $a = (0.15\pm0.07)\times 10^{-5} $ and $b = 0.25 \pm 0.03$.

The plot makes clear that increasing the cluster density by two orders of magnitude leads to $\sim 2.5\times$ more mergers. Moreover, it further highlights the role of primordial binaries, showing that clusters with a lower binary fraction have a probability $\sim 50\%$ smaller to develop a merger, at least in the case of $R_\ham = 1.75$ pc. 

\subsubsection{Merger rate for black hole binaries}

We define the cosmic merger rate following \cite{2020ApJ...898..152S,2021MNRAS.507.5224B}
\begin{align}
    \mathcal{R}(z) =& \frac{\der}{\der t_{\rm lb}(z)} \int_{0}^{z_{\rm max}}\psi_{\rm clus}(z')\frac{\der t_{\rm lb}(z)}{\der z'}\der z' \nonumber \\
    & \times \int_{Z_{\rm min}}^{Z_{\rm max}}\eta_{\rm GW}(Z)\mathcal{F}(z',z,Z)\der Z, 
    \label{rz}
\end{align}
where $t_{\rm lb}(z)$ is the lookback time at merger, $\psi_{\rm clus}(z')$ is the star cluster formation rate when the merging binary formed, $\eta_{\rm GW}(Z)$ is the merger efficiency at the metallicity $Z$, $\mathcal{F}(z',z,Z)$ is the number of mergers forming at redshift $z'$ and merging at redshift $z$ in environments with metallicity $Z$.
The adoption of Equation \ref{rz} enables us to compare \dragonii simulation results with those obtained for low-mass star clusters \citep{2020ApJ...898..152S}. Note that this procedure does not take into account possible effects related to the initial cluster mass function, which could indeed have an impact on the overall merger rate \citep[see e.g.][]{2020PhRvD.102l3016A}. Nonetheless, the similarity between the merger efficiency derived from \dragonii simulations and that obtained by \cite{2020ApJ...898..152S} for low-mass clusters suggests that it is possible to safely utilise the merger efficiency as a proxy of the overall number of mergers per unit mass in the whole range of possible cluster masses. This choice, although representing an approximation, permits us to avoid the inclusion of a cluster mass function in Equation \ref{rz} and all the related uncertainties, like the cluster mass function boundaries and functional form.

We adopt a cosmic star cluster formation rate in the form 
\begin{equation}
\psi_{\rm clus}(z) = \frac{0.01 (1+z)^{2.6}f_{\rm CFE} }{ 1 + \left[(1+z)/{3.2}\right]^{6.2}} ~\Ms {\rm yr}^{-1} {\rm Mpc}^{-3}, 
\label{eq:sfr}
\end{equation} 
i.e. we rescale the stellar star formation rate derived by \citep{2017ApJ...840...39M} by a factor $f_{\rm CFE}$, which represents the cluster formation efficiency, i.e. the fraction of star formation that goes into bound clusters. Although uncertain, observations and models suggest that the cluster formation efficiency (CFE) can be as large as $f_{\rm CFE,YC} = 0.3$ for young clusters \citep{2021Symm...13.1678M} and $f_{\rm CFE,GC} = 0.08\pm0.03$ \citep{2008MNRAS.390..759B} for globular clusters, regardless of the star formation history. In the following, we adopt both young and globular cluster CFE values to constrain the BBH merger rate in our simulations.

For dynamical mergers, it has been shown that the merger efficiency $\eta_{\rm GW}(Z)$ remains almost constant in the range $Z<10^{-3}$, and decreases roughly by an order of magnitude at solar values \citep{2020MNRAS.497.1043D,2020MNRAS.498..495D,2021MNRAS.507.3612R}. Since our models have all the same metallicity, $Z = 0.005$, to infer the merger rate we assume that the merger efficiency is constant at $Z<0.005$ and reduces by 10 times at larger metallicities \citep[see e.g. Figure 1 in][]{2020ApJ...898..152S}. Moreover, we factorise the function $F(z,z',Z) = p(Z,z') N(z,z')$, thus assuming that the number of mergers at redshift $z$ that formed at $z'$ is independent on the metallicity distribution. The $p(Z,z')$ term represents the cosmic fraction of clusters with metallicity in the ($Z$,$Z+dZ$) bin at redshift $z'$. 
We assume that the metallicity follows a log-normal distribution peaked at \citep{2017ApJ...840...39M}
\begin{equation}
    {\rm Log}\left\langle \frac{Z(z)}{{\rm~Z}_\odot}\right\rangle  = 0.153 - 0.074z^{1.34}, 
\end{equation}
with dispersion either $\sigma_Z = 0.2-0.5-0.8$ \citep{2020ApJ...892...13S,2021MNRAS.507.5224B,2021A&A...647A.153B,2021ApJ...910..152Z}.

Since all \dragonii models have the same metallicity, to infer the simulated merger rate we integrate Equation \ref{rz} under two assumptions, one conservative and one optimistic. In the conservative case, we consider only clusters with a metallicity $Z<0.005$ and assume that they have a similar merger rate efficiency \citep{2020ApJ...898..152S}. In the optimistic case, instead, we include in the integration also clusters with metallicity larger than the simulated one, reducing for metal-rich clusters the simulated merger efficiency by a factor 10, as expected from low-$N$ simulations of young clusters \citep{2019MNRAS.487.2947D}.

To compare with similar estimates in the literature, we first set $f_{\rm CFE} = 1$, i.e. that all stars form in star clusters, and calculate a merger rate of $\mathcal{R} = 27 \yrgpc$, in broad agreement with the rate inferred for low-mass star clusters ($N=10^2-5\times 10^4\Ms$) \citep{2020MNRAS.498..495D,2021MNRAS.507.3612R,2020ApJ...898..152S} and semi-analytic models of young and globular clusters \citep[see e.g.][]{2021Symm...13.1678M}.

\begin{figure}
    \centering
    \includegraphics[width=\columnwidth]{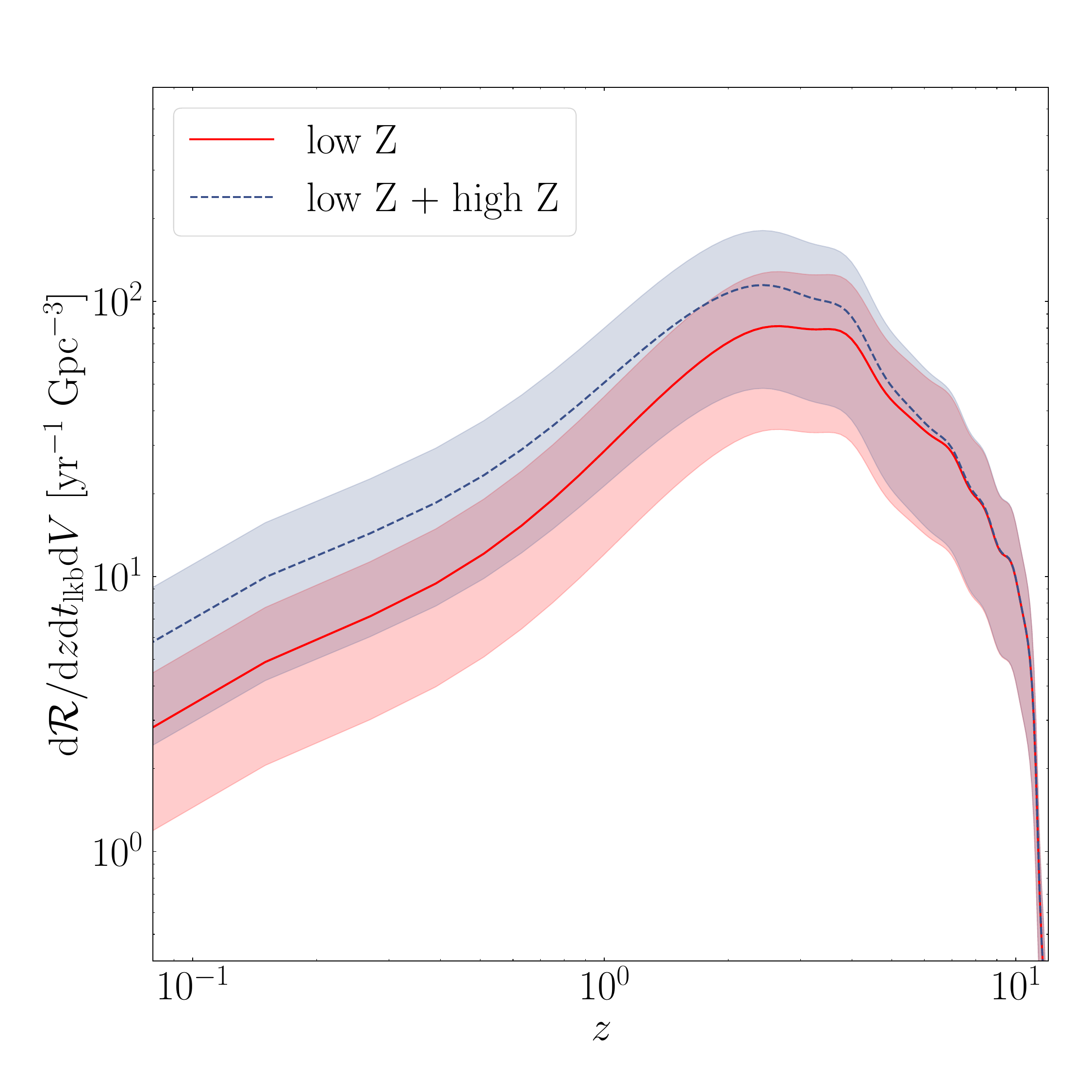}
    \caption{Merger rate density for all simulations in our runs, assuming either metal-poor only (red straight line) or both metal-poor and rich (black dashed line) clusters. The shaded area embrace two limiting values of the cluster formation efficiency, with the upper limit corresponding to $f_{\rm CFE}=0.3$ and the lower one corresponding to $f_{\rm CFE} = 0.08$.}
    \label{fig:GWrate}
\end{figure}

A more reliable estimate of the merger rate is shown in Figure \ref{fig:GWrate} for both the conservative and optimistic cases, and assuming different values of the cluster formation efficiency, $f_{\rm CFE}=0.08-0.3$. 
As shown in the plot, we find a simulated merger rate of $\mathcal{R}_{\rm GW} = (12\pm7)$ at redshift $z=0.2$. At the same redshift, the BBH merger rate inferred by the LVK is  $\mathcal{R}_{\rm LVK}=17.9-44 \yrgpc$ \citep{2021arXiv211103606T,2021arXiv211103634T}.

\begin{table}
    \centering
    \begin{tabular}{c|c}
    \hline
    \hline
        Source type & $\mathcal{R}_{\rm loc}$\\
                & $\yrgpc$\\
    \hline
        BBH   & $5-19$ \\
        NS-BH & $0.027-8.7$ \\
        WD-BH & $3.8\times 10^{-4}-2.3$ \\
    \hline
    \end{tabular}
    \caption{Volumetric merger rate calculated in the local Universe for different GW sources in \dragonii clusters. }
    \label{tab:rate}
\end{table}

\subsubsection{Merger rate for exotic mergers}
In \dragonii simulations we find 3 elusive mergers: one WD-BH and two NS-BH mergers.  Despite they are evidently too scarce to allow a statistical treatment, we can exploit them to attempt a rough, order of magnitude, estimate of the merger rates for these two classes of GW sources assembled in star clusters as:
\begin{equation}
    R_{\rm xBH}(<D) = \frac{N_{\rm x}}{M_{\rm sim}} f_{x} \delta M_{g*} N(<D) t_{\rm rel}^{-1},
    \label{eq:exrate}
\end{equation}
where $M_{g*}$ is the galaxy stellar mass, $\delta=0.001-0.01$ is the fraction of galaxy mass made up by star clusters \citep{2009MNRAS.392L...1S,2010MNRAS.406.1967G,2013ApJ...772...82H,2014MNRAS.444.3738A,2014ApJ...785...71G,2015MNRAS.453.3278W}, $f_{x}$ is the fraction of clusters with a given property (e.g. age within a certain range), 
$t_{\rm rel}$ is the cluster relaxation time, and $N(<D)$ is the number of MW equivalent galaxies within a given cosmological distance $D$ \citep{2010CQGra..27q3001A}
\begin{equation}
    N(<D) = 4\pi/3 (2.26)^{-3} \left(\frac{D}{{\rm Mpc}}\right)^3 \frac{\rho_g}{{\rm Mpc^{-3}}},
\end{equation}
where $\rho_g = 0.0116$ Mpc$^{-3}$ is the galaxy number density in the local Universe \citep{2008ApJ...675.1459K}.
Moreover, we consider typical relaxation times of either globular clusters, $t_{\rm rel} = 10^9$ yr \citep{2010arXiv1012.3224H}, or massive and relatively young clusters in the Small Magellanic Cloud (SMC), $t_{\rm rel} = 3.2\times 10^7$ yr \citep{2021MNRAS.507.3312G}. 

Note that the relaxation time of Galactic clusters is inferred from their present time properties. Depending on the amount of mass lost and the level of cluster expansion, it could be possible that the initial relaxation time was relatively shorter and therefore the number of {\it dynamically old} globular clusters is larger than what we see at present. In these regards, note that the relaxation time of SMC clusters, which are generally younger than a few Gyr, is sensibly smaller compared to Milky Way globulars, possibly because relaxation processes did not have time to sufficiently influence the cluster dynamics.

In the following calculations, we consider Milky Way-like galaxies only, $M_{g*} = 6\times 10^{10}\Ms$ \citep{2015ApJ...806...96L} located within $D = 1$ Gpc. In the Milky Way, there are only $\sim 4$ out of 155 globular clusters with an age larger than 1 relaxation time, whilst around half of clusters in the SMC satisfy this requirement, thus $f_{x} \sim 0.025 - 0.5$. 

This implies a frequency rate for WD-BH mergers in the local Universe of $R_{\rm WDBH} = (1.8\times10^{-3} - 10.8)$ yr$^{-1}$, corresponding to a volumetric merger rate $\mathcal{R}_{\rm WDBH} = RV_{\rm com}^{-1}(1{\rm~Gpc}) = (3.8\times10^{-4}-2.3)\yrgpc$.

In the case of NS-BH mergers, instead, the event occurs over a timescale of $t_\gw = (0.04-0.5)t_{\rm rel}$. The fraction of cluster with an age longer than $t_\gw$ is $f_x\sim 0.94$ for clusters in both the Milky Way and the SMC, the resulting frequency rate for NS-BH mergers is $R_{\rm NSBH} = (0.13-40.7)$ yr$^{-1}$, which implies a volumetric merger rate of $\mathcal{R}_{\rm NSBH} = (0.027-8.7) \yrgpc$. 

\subsection{Multimessenger sources: prospects for LISA detection}

Over the next decade, the network of ground-based detectors will be complemented by LISA, possibly the first space-borne low-frequency detector. LISA will be able to possibly catch the GW echoes of merging stellar BHs, IMBHs, and nearby WD and NS binaries. While we postpone a detailed discussion about BBHs forming in young massive clusters detectable with LISA to a forthcoming paper, we focus in the following on the handful exotic mergers that develop in our \dragonii models.

Let us consider the case of a WD-BH merger. We have shown in Section \ref{sec:exo} that such a source could appear as an X-ray binary and give rise to a TDE once the WD approaches too closely the BH. Assuming that the binary evolves solely via GW emission, and adopting the \cite{1963PhRv..131..435P} formalism to evolve the binary until the merger, we find that around 6 months prior to the merger, the WD will overfill the Roche lobe and start the X-ray binary phase. 
At disruption, the frequency of the associated GW emission is given by \citep{1995ApJ...445L...7H,2004ApJ...615..855K,2009ApJ...695..404R,2013MNRAS.434.2948D,2020MNRAS.495.1061F}
\begin{align}
    f_{\rm GW} \simeq& 0.09{\rm Hz} \left(1+\frac{M_{\rm WD}}{M_{\rm BH}}\right) \times \nonumber \\
                     & \times \left(\frac{M_{\rm WD}}{0.6\Ms}\right)^{1/2}\left(\frac{R_{\rm WD}}{10^4{\rm km}}\right)^{-3/2} = 0.13{\rm Hz},
\end{align}
where we have assumed $R_{\rm WD} = 10^4$ km. Note that an eccentricity between 0 and 1 would affect $f_{\rm GW}$ by less than $20\%$ \citep{2020MNRAS.495.1061F}. The amplitude of the emitted signal at disruption will be \citep{2019CQGra..36j5011R,2020MNRAS.495.1061F}
\begin{align}
    h_c\simeq & 2\times10^{-20}\left(\frac{T_{\rm obs}}{4{\rm yr}}\right)^{1/2} \left(\frac{D_L}{10{\rm Mpc}}\right)^{-1}\left(\frac{M_{\rm BH}}{10\Ms}\right)^{0.66}\times \nonumber \\
    & \times \left(\frac{M_{\rm WD}}{0.6\Ms}\right)^{1.58}\left(\frac{R_{\rm WD}}{10^4{\rm km}}\right)^{-1.75} \simeq 10^{-19}.
\end{align}    
Since the WD will disrupt completely as crossing its Roche limit, the associated GW emission will appear as a burst \citep{2009ApJ...695..404R}. For such source, the corresponding signal-to-noise ratio (S/N) for LISA can be written as \citep[see e.g.][]{2019CQGra..36j5011R}
\begin{equation}
    \left({\rm S}/{\rm N}\right) = f^{2/3}\frac{h_c}{S_c} = 1.2\left(\frac{D_L}{10{\rm Mpc}}\right)^{-1},
\end{equation}
where $S_c$ is the detector sensitivity curve in terms of characteristic strain \citep{2019CQGra..36j5011R} and we have exploited the intrinsic dependence on the measurable GW strain and the source luminosity distance $D$. 

If the merger occurs inside the Milky Way, i.e. at $D < 0.1$ Mpc, it would appear as a loud source in LISA, with (S/N)$> 120$. More in general, the maximum distance at which LISA could detect such merger with a minimum signal-to-noise ratio of (S/N)$>8(15)$ is $D < 1.5{\rm ~Mpc}$($0.7{\rm ~Mpc}$).

Note that the Andromeda galaxy is $\sim 0.7-0.8$ Mpc away from us, therefore to roughly estimate the probability for a closeby WD-BH merger we can replace in Equation \ref{eq:exrate} $N(<D) = 2$ and find an upper limit to the local merger rate of closeby WD-BH mergers of $R_{\rm WDBH,close} < (8.4\times 10^{-10} - 5.1\times 10^{-6})$ yr$^{-1}$.

\subsection{The pair-instability supernova rate for massive star clusters: perspectives for detection via magnitude limited surveys}
\label{sec:pisnerate}

The onset of IMBH formation and the development of BBH mergers depend intrinsically on the cluster radius and initial density, the amount of stars initially in a binary, and the stellar evolution recipes adopted -- e.g. BH matter accretion efficiency, the physics of PISNe and PPISNe. 

In these regards, the fact that PISNe are rare events for which a smoking gun has not been observed yet \citep[e.g.][]{2007Natur.450..390W,2009Natur.462..624G,2017NatAs...1..713T,2019ApJ...881...87G, 2022ApJ...938...57W}, offers us the possibility to use this physical process as a diagnostic quantity in \dragonii models. 
In practice, we can infer the PISN rate in \dragonii simulations and compare such rate with current observation limits to explore whether our simulations produce unrealistically large PISN frequency rates.
As described in paper AS-II, in \dragonii models PISNe develop either in single stars or in stellar merger products, provided that their core Helium reaches a mass in the range $(64-130)\Ms$. This offer us a unique possibility to explore the impact of PISNe in star clusters, taking simultaneously into account the impact of stellar mergers in the overall population of PISN progenitors. According to the adopted stellar evolution, in a simple stellar population only stars heavier than $m_{\rm ZAMS}\geq m_{\rm PISN} = 150\Ms$ could undergo a PISN event, i.e. larger than the maximum stellar mass adopted for the initial mass function. Instead, in \dragonii models we find 23 stars that undergo supernova. All these stars are either in a primordial binary or are captured in a binary before the explosion and undergo one or more stellar merger and accretion events that bring the star mass above $m_{\rm PISN}$. Typical masses for \dragonii PISN progenitors are in the range $(150-282)\Ms$.

The simulated PISN efficiency can be defined similarly to the compact object merger rate, i.e. $\eta_{\rm PISN} = N_{\rm PISN}/M_{\rm sim} = 6.2\times 10^{-6}\Ms^{-1}$. 

To calculate the PISN rate, we follow the approach adopted by \cite{2020MNRAS.499.5941D}. Firstly, we assume that the Ni mass of the massive star that goes off as a PISN can be calculated via the following equation:
\begin{equation}
    {\rm Log} \left(M_{\rm Ni}/\Ms\right) = r \left(M_{\rm He, f}/\Ms\right)^s + t,
\end{equation}
where $r = -5.02\times 10^4$, $s = -2.159$, and $t = 2.887$ \citep{2002ApJ...567..532H,2020MNRAS.499.5941D}, and $M_{\rm He,f}$ is the final mass of the star He core. The Ni mass is used to infer the peak bolometric magnitude exploiting an Arnett-like relation \citep{1982ApJ...253..785A}
\begin{equation}
    \Upsilon_{\rm bol, Ni}^{\gw} = -19.2-2.5{\rm Log}\left(M_{\rm Ni}/0.6\Ms\right),
\end{equation}
which can be converted into an apparent bolometric magnitude via the Pogson's relation
\begin{equation}
    \mu_{\rm bol}^{\gw} = \Upsilon_{\rm bol, Ni}^{\gw} + 5{\rm Log}(D_L / 10{\rm pc}),
\end{equation}
being $D_L$ the luminosity distance. To simplify the calculations, we adopt for the He mass, which is the main ingredient to calculate the Ni mass, an average value of $M_{\rm He,f} = 90.4\Ms$ as extracted from our models. 

The value of $\mu_{\rm bol}^{\gw}$ is used to calculate whether a PISN can be detected in a magnitude limited survey. Assuming to have a population of PISNe with apparent magnitude distributed according to a Gaussian around $\mu_{\rm bol}^{\gw}$ and a magnitude detection threshold $\mu_{\rm lim}$, we define the fraction of detectable sources as 
\begin{equation}
    f_{\rm GSS} = 0.5\left[1+{\rm erf}\left(\frac{\mu_{\rm lim}-\mu_{\rm bol}^{\gw}}{\sqrt{2}\sigma_\mu}\right)\right]    ,
\end{equation}
where we adopted $\sigma_\mu = 0.2$\footnote{We verified that varying $\sigma_\mu$ in the range $0.1-0.3$ has little effect on our results.}.

\begin{figure}
    \centering
    \includegraphics[width=\columnwidth]{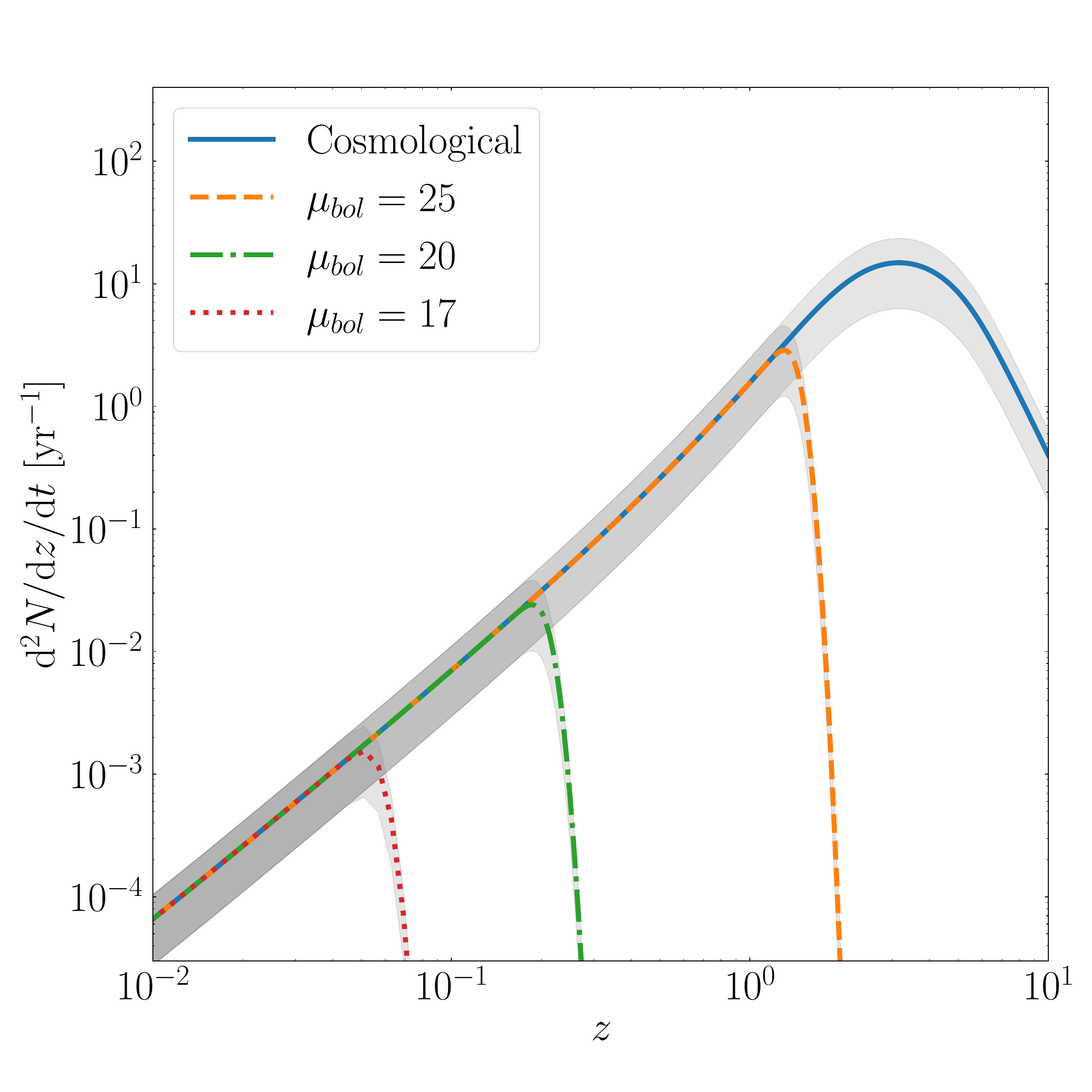}
    \caption{PISNe cosmological rate (blue straight line) and detection rate assuming a magnitude limited survey at different threshold values, namely $m_{\rm bol}=17$ (red dotted line), $20$ (green dashed-dotted line), and $25$ (orange dashed line), assuming a cluster formation efficiency of $f_{\rm CFE} = 1$ and assuming that only clusters with a metallicity $Z \lesssim 0.008$ can host stars undergoing PISN.}
    \label{fig:pisnrate}
\end{figure}

The PISN rate as a function of the redshift can thus be evaluated as:
\begin{equation}
    \mathcal{R}_{\rm PISN}(z) = \int_{z_1}^{z_2} \frac{dV}{dz} \psi(z) \eta_{\rm PISN} f_{\rm GSS}(z) f_{\rm Z}(z) dz,
    \label{eq:pisnrate}
\end{equation}
where $dV/dz$ is the comoving volume element and $\psi(z)$ is the cosmic star formation rate, for which we assume the cosmic star formation history in Equation \ref{eq:sfr} \citep{2017ApJ...840...39M} and the same limits for $f_{\rm CFE}$ described in Section \ref{sec:mergerate}, and that only stars with a metallicity $Z \leq 0.008$ undergo PISNe \citep{2017MNRAS.470.4739S}. Figure \ref{fig:pisnrate} shows the PISNe rate for the intrinsic cosmic population and assuming different detection threshold in magnitude limited surveys, namely $\mu_{\rm bol} = 17,~20,~25$. Note that these threshold roughly corresponds to the typical maximum detectable magnitude of already completed, like the Sloan Digital Sky Survey (SDSS\footnote{SDSS home: \url{http://www.sdss.org}}) or the Palomar Transient Factory (PTF\footnote{PTF home: \url{http://www.ptf.caltech.edu}}), ongoing, e.g. the Dark Energy Survey (DES\footnote{DES home: \url{http://www.darkenergysurvey.org}}), and future surveys, like the Large Synoptic Survey Telescope (LSST\footnote{LSST home: \url{http://www.lsst.org}}), the Zwicky Transient Facility (ZTF\footnote{ZTF home: \url{https://www.ztf.caltech.edu}}), or the EUCLID mission\footnote{EUCLID home: \url{https://sci.esa.int/web/euclid}} \citep{2022A&A...666A.157M}.
From Figure \ref{fig:pisnrate} we see that only future surveys ($\mu_{\rm bol} \geq 25$) will be able to probe the cosmological properties of PISNe, whilst current surveys could in principle place constraints on PISNe within a redshift $z<0.3$.

Integrating Equation \ref{eq:pisnrate} over the redshift returns the number of detected sources per year. The possible number of PISN detections per year for different values of the limiting bolometric magnitude, $\mu_{\rm bol}$, and the cluster formation efficiency, $f_{\rm CFE}$, is summarized in Table \ref{tab:pisnrate}. From the table is clear that the detection of PISNe from star clusters is still highly unlikely in completed and ongoing surveys, but it could lead to $\sim 8$ detections per year with the next generation of detectors. Comparing future PISNe detections with numerical models could have a twofold aim. On the one hand, it will permit us to shed light on the actual contribution of massive stars in dense clusters to the overall population of PISNe. On the other hand, it will provide us with an useful term of comparison to determine the reliability of cluster simulations.

\begin{table}
    \centering
    \caption{Cosmological and detectable PISNe occurrence rate. Col. 1: cluster formation efficiency. Col. 2-5: number of PISNe per year for the cosmological population, and for magnitude limited surveys with different treshold magnitudes.}
    \begin{tabular}{c|cccc}
     \hline\hline
     $f_{\rm CFE}$ & \multicolumn{4}{c}{No. of detections per 1 year}\\
     & Cosmological & $m_{\rm bol} = 25$ & $m_{\rm bol} = 20$ & $m_{\rm bol} = 17$ \\ 
     \hline
     $0.08$&  $  24 $& $ 0.7$ & $0.001$ &  $ 1.8\times 10^{-5}$ \\ 
     $0.3$ &  $  89 $& $ 2.6$ & $0.004$ &  $ 6.8\times 10^{-5}$ \\
     $1$   &  $ 297 $& $ 8.8$ & $0.012$ &  $ 2.3\times 10^{-4}$ \\
    \hline
    \end{tabular}
    \label{tab:pisnrate}
\end{table}

\section{Conclusions}
\label{sec:end}
In this paper we have presented and discussed the properties of compact binary mergers and PISNe in the \dragonii simulations, a suite of direct $N$-body models representing star clusters with up to 1 million stars and a relatively large ($10\%-33\%$) binary fraction. Our main results can be summarised as follows:
\begin{itemize}
    \item We find a population of 75 BBH, 2 NS-BH, and 1 WD-BH mergers. Among them, 4 BBHs avoid merger when GW recoils are enabled. Mergers occurring inside the cluster make-up the $\gtrsim 40\%$ of the whole population and are mostly due to mergers formed via dynamical interactions (dynamical mergers). The population of ejected mergers, which merge outside the parent cluster, are equally contributed by mergers formed dynamically and from primordial binaries (primordial mergers). Typically, in-cluster mergers have primaries with masses $m_{\rm BH,1} > 30\Ms$ and companion in the $m_{\rm BH,2} = 30-50\Ms$ mass range, whilst ejected mergers involve lighter primaries, $m_{\rm BH,1} < 40\Ms$, and are characterised by fairly large mass ratios, $q > 0.6$;
    \item Mergers forming from primordial binaries are characterised by large mass ratios and component masses clearly smaller than those formed dynamically. Among dynamical mergers, the most massive ones are those in which at least one component had an ancestor in a primordial binary;
    \item BBH mergers are characterised by a delay time that nicely distribute around a value of $10-30$ cluster relaxation time. This highlights the fact that the processes that trigger BBH formation and merger are intrinsically related to the cluster dynamical evolution;
    \item The population of mergers forming from dynamical interactions or primordial binaries is clearly distinguishable from the residual eccentricity of the binary as it enters in the typical frequency bands of GW detectors, i.e. $f = 0.001-100$ Hz. We find that practically all primordial binaries are circular at merger, this implying that primordial binaries merge before dynamics can have an impact on their evolution, whilst around $20-40-5\%$ of mergers preserve an eccentricity $e > 0.1$ when entering the LISA-DECIGO-LIGO bands. All mergers with $e > 0.1$ in the 0.05-1 Hz and 1-10 Hz bands occur inside the cluster, whilst half of eccentric mergers in the mHz band are ejected. This hints at the possibility to distinguish the formation history of a BBH merger from the frequency band in which it is observed;
    \item We identify three exotic mergers in our sample: a WD-BH binary formed dynamically and two NS-BH mergers, one formed dynamically and the other from a primoardial binary. A WD-BH merger that forms after 4 cluster relaxation time and it is triggered by chaotic interactions that increase the eccentricity up to an extremal value of $e = 0.99994930$. Once the WD approaches sufficiently close the BH, this type of sources could appear as an ultraluminous X-ray sources and, ultimately, be a source detectable by LISA if it occurs within 700 kpc from us, i.e. within the distance between the Milky Way and Andromeda. The dynamical NS-BH binary is characterised by a chirp mass $\mathcal{M} = 3.4\Ms$, larger than what predicted by the isolated stellar evolution scenario, and preserve an eccentricity of $e= 0.9974(0.21)$ when crossing a frequency of $f = 0.5(1)$ Hz, thus future observations with ET could help probing the population of closeby, dynamically formed, NS-BH mergers. The primordial NS-BH binary is not affected by dynamics at all, thus they can be mistaken for a merger occurring in isolation. This highlights the importance of star clusters with a large binary fraction as contributors of the isolated scenario of compact binary mergers. None of the NS-BH mergers are expected to release EM emission, unless the BHs have a spin $\chi > 0.9$;
    \item We find that comparing the remnant mass and spin of BBH mergers could help untangling their origin. Using a model based on stellar evolution theories, we show that primordial binary mergers are characterised by remnant masses systematically smaller and effective spin parameters systematically larger than dynamical mergers;
    \item We derive a BBH merger efficiency of $\sim 2\times 10^{-5} \Ms^{-1}$, comparable with the value estimated for low-mass star clusters. Interestingly, we find that the merger efficiency depends on the star cluster properties. Decreasing the binary fraction by a factor $4$, for example, leads to a decrease of the merger efficiency by a factor $\sim 2$. Moreover, the merger efficiency increases with the cluster density following a power-law with slope $\sim 0.25$. We adopt a series of cosmologically motivated assumptions for the cosmic star formation history, and use them to infer a merger rate density at redshift $z < 0.2$ of $\mathcal{R} = 5-19 ~(0.027-8.7)~(3.8\times10^{-4} - 2.3) \yrgpc$ for BBHs(WD-BH)(NS-BH) mergers, respectively. We predict that, in a 4 yr-long mission, LISA could detect $N_{\rm BBH} = 12\pm7$($5\pm3$) BBH mergers (IMRIs) and can identify the WD-BH merger with a signal-to-noise ratio SNR$ > 8$($15$) if it occurs within $D_L < 1.5$($0.7$) Mpc from us. 
    \item We retrieve the cosmic frequency rate of PISNe, in order to explore the reliability of our simulations on the one hand, and to make predictions for PISNe detection from star clusters on the other hand. We find that future surveys with a limiting magnitude of $m_{\rm bol} = 25$ could detect $N_{\rm PISN} = 0.7-8.8$ PISNe per year. Comparing these estimates with future surveys could help placing constraints on the population of massive stars in dense star clusters. 
    \end{itemize}

The \dragonii clusters represent a further step forward in the modelling of young and intermediate-age star clusters, providing the first suite of simulations that models clusters with both $N>120,000$ stars (up to $10^6$), a high binary fraction (up to $33\%$), and an initial density of $\rho = (1.2\times 10^4-1.6\times10^6)\Ms$ pc$^{-3}$. These simulations complement the vast literature of $N$-body simulations of lower-mass and lower density star clusters \citep[e.g.][]{2019MNRAS.487.2947D,2021MNRAS.501.5257R,2021MNRAS.507.3612R,2021MNRAS.500.3002B,2022MNRAS.512..884R,2022A&A...665A..20B}, and provide the largest catalogue of BH mergers obtained in direct $N$-body simulations of metal-poor, dense and massive young clusters.

\section*{Acknowledgements}
The authors thank the referee for their constructive and helful report. The authors warmly thank Agostino Leveque for their help and assistance in using their implementation of the \mcl code, and Giuliano Iorio, Sara Rastello, and Michela Mapelli for useful comments and discussion. 

This work benefited of the support from the Volkswagen Foundation Trilateral Partnership through project No.~97778 ``Dynamical Mechanisms of Accretion in Galactic Nuclei'' and the Deutsche Forschungsgemeinschaft (DFG, German Research Foundation) -- Project-ID 138713538 -- SFB 881 ``The Milky Way System''), and by the COST Action CA16104 ``GWverse''. The authors gratefully acknowledge the Gauss Centre for Supercomputing e.V. for funding this project by providing computing time through the John von Neumann Institute for Computing (NIC) on the GCS Supercomputer JUWELS Booster at Jülich Supercomputing Centre (JSC). Data analysis and part of the runs were conducted on the GRACE-BH HPC workstation, funded by the European Union's under the research project GRACE-BH.

MAS acknowledges funding from the European Union’s Horizon 2020 research and innovation programme under the Marie Skłodowska-Curie grant agreement No.~101025436 (project GRACE-BH, PI: Manuel Arca Sedda). 

AWHK is a fellow of the International Max Planck Research School for Astronomy and Cosmic Physics at the University of Heidelberg (IMPRS-HD).
The work of PB was supported by the Volkswagen Foundation under the special stipend No.~9B870.

PB acknowledge the support within the grant No.~AP14869395 of the Science Committee of the Ministry of Science and Higher Education of Kazakhstan ("Triune model of Galactic center dynamical evolution on cosmological time scale"). 
The work of PB was supported under the special program of the NRF of Ukraine Leading and Young Scientists Research Support - "Astrophysical Relativistic Galactic Objects (ARGO): life cycle of active nucleus", No.~2020.02/0346.

RS thanks Max Planck Institute for Astrophysics (Thorsten Naab) for hospitality during many visits

MG was partially supported by the Polish National Science Center (NCN) through the grant No. 2021/41/B/ST9/01191.

FPR acknowledges the support by the European Research Council via ERC Consolidator Grant KETJU (no. 818930).

TN acknowledges the support of the Deutsche Forschungsgemeinschaft (DFG, German Research Foundation) under Germany’s Excellence Strategy - EXC-2094 - 390783311 of the DFG Cluster of Excellence "ORIGINS”.
\section*{Data Availability}
The data from the runs of these simulations and their initial models
will be made available upon reasonable request by the corresponding author. 
The \textsc{Nbody6++GPU} code is publicly available\footnote{\url{https://github.com/nbody6ppgpu/Nbody6PPGPU-beijing}}. The \textsc{McLuster} version used in this work will soon be available. A similar version is described in \cite{2022MNRAS.514.5739L}.


\bibliographystyle{mnras}
\bibliography{example} 

\bsp	
\label{lastpage}
\end{document}